\documentclass[fleqn]{2017SCGE}
\usepackage[T1]{fontenc}	
\usepackage{footmisc}			
\usepackage{graphicx,subfigure,tabularx}
\usepackage[all]{hypcap}
\usepackage{tcolorbox,color}
\usepackage{amsfonts,amsmath,amstext,amssymb,mathrsfs,gensymb}
\usepackage{makecell,multirow}
\usepackage{hyperref}
\begin{document}
\ensubject{subject}

\ArticleType{Article} 
\Year{2018}
\Month{October}
\Vol{??}
\No{?}
\DOI{?}
\ArtNo{000000}
\ReceiveDate{September 30, 2018}
\AcceptDate{??, 2018}

\title{A high-resolution self-consistent whole sky foreground model}{A high-resolution self-consistent whole sky foreground model}

\author[1,2]{Qizhi Huang}{}%
\author[1]{Fengquan Wu}{}
\author[1,3,4]{Xuelei Chen}{{xuelei@cosmology.bao.ac.cn}}

\AuthorMark{Huang Q Z}

\AuthorCitation{Huang Q Z, Wu F Q \& Chen X L}

\address[1]{Key Laboratory of Computational Astrophysics, National Astronomical Observatories, Chinese Academy of Sciences, Beijing 100101, China}
\address[2]{Universit\'e Paris-Sud, LAL, UMR 8607, F-91898 Orsay Cedex, France $\&$ CNRS/IN2P3, F-91405 Orsay, France}
\address[3]{School of Astronomy and Space Science, University of Chinese Academy of Sciences, Beijing 100049, China}
\address[4]{Center for High Energy Physics, Peking University, Beijing 100871, China}



\abstract{
The neutral hydrogen 21cm line is potentially a very powerful probe of the observable universe, and a number of on-going 
experiments are trying to detect it at cosmological distances. However, the presence of strong foreground radiations 
such as the galactic synchrotron radiation, galactic free-free emission and extragalactic radio sources 
make it a very challenging task. For the design of 21cm experiments and analysis of their data, simulation is an essential tool, 
and good sky foreground model is needed. With existing data the whole sky maps are available only in low angular resolutions 
or for limited patches of sky, which is inadequate in the simulation of these new 21cm experiments. 
In this paper, we present the method of constructing a high resolution self-consistent sky model at low frequencies, 
which incorporates both diffuse foreground and point sources. 
Our diffuse map is constructed by generating physical foreground components including 
the galactic synchrotron emission and galactic free-free emission. The point source sample is generated using the 
actual data from the  NRAO VLA Sky Survey (NVSS) and the Sydney University Molonglo Sky 
Survey (SUMSS) where they are available and complete in flux limit, and mock point sources 
according to statistical distributions. The entire model is made self-consistent by removing the integrated flux of 
the point sources from the diffuse map so that this part of radiation is not double counted. 
We show that with the point sources added, a significant angular power is introduced in the mock sky map, 
which may be important for foreground subtraction simulations. 
Our sky maps and point source catalogues are available to download. 
}



\maketitle


\begin{multicols}{2}

\section{Introduction}

Tomographic observation of the redshifted 21cm signal from neutral hydrogen will allow us to study the 
Epoch of Reionization (EoR) \cite{1997ApJ...475..429M, 2003ApJ...596....1C} and the Cosmic Dawn \cite{2015MNRAS.447.1806G}, 
to trace the evolution of the large scale structure of our universe and constrain the 
cosmological parameters including the equation of state of dark energy \cite{2008PhRvL.100i1303C, 2008PhRvD..78b3529M}, 
and even to probe the cosmic dark age \cite{2004PhRvL..92u1301L}. 
A number of telescopes have been developed to detect the cosmological 21cm signal, including the EoR experiments such as
LOFAR \cite{2013A&A...556A...2V}, MWA \cite{2013PASA...30....7T}, PAPER \cite{2010AJ....139.1468P},  
and HERA \cite{2017PASP..129d5001D};  
and the dark energy experiments such as Tianlai \cite{2012IJMPS..12..256C} and CHIME \cite{2014SPIE.9145E..22B};   
and the next generation telescopes such as SKA \cite{2013arXiv1311.4288H} which are expected to provide considerably 
more accurate power spectrum and directly map the large scale structures. 
	
However, detecting the cosmological 21cm signal is very difficult due to the presence of foregrounds 
which are several orders of magnitude stronger. At low frequencies the main astrophysical foregrounds include the 
galactic synchrotron emission which arises from the relativistic electrons moving in the galactic magnetic field, 
the galactic free-free emission which is produced by the scattering of free electrons with diffuse warm ionized gas, 
strong radio sources such as supernova remnants (SNRs), and extragalactic radio sources such as radio-loud galaxies, 
quasars and BL Lac objects \cite{2008MNRAS.389.1319J,1999A&A...345..380S}. 
In order to study the foreground removal methods and evaluate its impacts on the 21cm experiments by numerical simulation, 
mock sky images with foregrounds are needed.
	
Various studies on the foregrounds have been carried out for the 21cm experiments 
\cite{2004MNRAS.355.1053D, 2008MNRAS.388..247D, 2012MNRAS.419.3491L, 2017MNRAS.464.3486Z, 2014ApJ...781...57S, 2015PhRvD..91h3514S}. 
To investigate how the different foreground subtraction techniques work, or to make forecast of the experiments, 
simulations with foreground sky models are needed. 
The Global Sky Model (GSM; \cite{2008MNRAS.388..247D}) is widely used for this purpose, 
with a recent updated version \cite{2017MNRAS.464.3486Z}. 
We denote the original and updated GSM models as  GSM2008 and GSM2016 respectively. 
The foreground map in the GSM model is based on the method of Principal Component Analysis (PCA). 
GSM uses several whole sky or large area sky maps, especially the Haslam 
map \cite{1982AandAS...47....1H} at 408 MHz as input data, to extract the principal components in the 
pixel-frequency space, to generate maps at other frequencies. 
GSM2008 applies the PCA once on the whole pixel-frequency data set, 
while GSM2016 applies the PCA  iteratively on the outputs of each iterations. 
However, owing to the limitation of the available whole sky surveys (e.g. the Haslam map 
has an angular resolution of $0.85^\circ$), the angular resolutions of these models are limited at low frequencies.

A high resolution whole sky model is provided in the 21cm simulation 
software package CORA \cite{2014ApJ...781...57S, 2015PhRvD..91h3514S}.
It uses the 2003 version of the Haslam map \footnote{\url{https://lambda.gsfc.nasa.gov/product/foreground/haslam_408.cfm}} 
as the base map, with the spectrum index given by GSM, plus a randomly generated variation in the spectrum index. 
To include structures on the small scales, CORA added bright point sources drawn from the matched ones in the
VLSS \cite{2007AJ....134.1245C} and NVSS \cite{1998AJ....115.1693C} surveys, plus a mock sample of randomly generated
point sources according to the distribution given in Ref.\cite{2002ApJ...564..576D}, and finally, 
based on Ref.\cite{2005ApJ...625..575S},
very faint point sources are taken into account as a whole in the small scale angular power spectrum.
There are also other high resolution models, e.g. the Tiered Radio Extragalactic Continuum Simulation 
(T-RECS; \cite{2018arXiv180505222B}), but only a limited patches of the sky are covered.

In this study, we construct a high-resolution whole sky foreground model 
by modeling the sky as the sum of a diffuse component and a point source component. 
For the diffuse component, our model differs from the GSM model
in that instead of using the purely mathematical approach of PCA in GSM (2008 and 2016), 
we fit the spectral index, varying both with frequency and angular direction using the multi-wavelength data. 
Such model is less general than the PCA model, but may have some advantages in certain applications, 
e.g. employing data from other wave bands to improve the foreground model, or to make more extreme extrapolation. 
We also use the updated Haslam map which is source-subtracted and 
de-striped \cite{2015MNRAS.451.4311R}\footnote{\url{http://www.jb.man.ac.uk/research/cosmos/haslam_map}} as our basis.
We achieve a higher angular resolution by adding point sources to the low resolution maps. This procedure is 
possible because point sources, i.e. sources unresolved up to the angular resolution of the survey, 
have been detected in surveys of higher angular resolutions, further, their statistical distribution is known or can be inferred. 
However, these point sources are not detected individually in surveys with low angular resolutions, 
but are merged into the diffuse foreground. 
Developed independently, our procedure differs from that of CORA in detail, for example 
in the choice of statistical model, and the placement of the points--CORA assumed a uniformly random distribution of points, 
whereas we consider angular clustering of the radio sources and use Rayleigh-L\'evy random walk to place the 
mock point sources. However, we found that our results are similar to that obtained using
 CORA.
	
The remainder of this paper is organized as follows. In Section 2, we outline the construction of our foreground model. 
In Sections 3 and 4, we present our method of constructing the diffuse map and point source contribution respectively. 
Finally, the resulting total sky map is presented in Section 5 and discussed.


\section{Foreground Model}

In an astronomical observation, a source of radiation can be classified either as an extended source or a point source, where
the latter is unresolved up to the limit of angular resolution. This classification depends on the angular resolution 
of the observation: at higher resolution some point sources in low resolution surveys will turn out to be extended ones. 

In our sky model, we consider two types of contributions: a whole sky diffuse foreground which include all extended sources,
and the point sources. The sky intensity for an angular resolution $\Theta$ is then given by 
\begin{eqnarray}
	I_{\Theta}(\vec{n},\nu) = I^f_{\Theta}(\vec{n},\nu) + \sum_{p \in \textbf{M}(\Theta)} I^p_{\Theta}(\vec{n},\nu) 
	\label{eq:skymodel1}
\end{eqnarray}
where $\vec{n}$ denotes the direction of the sky, $\nu$ denotes the frequency, $I^f_{\Theta}$ denotes
 the whole sky diffuse foreground, 
$\textbf{M}_{\Theta}$ denotes the set of point sources at resolution $\Theta$, and $I^p_{\Theta}$ denotes the 
radiation from point source $p$ convolved with a beam of angular resolution $\Theta$. 
The observed diffuse foreground map and catalog of point sources are available only at some frequencies, and in most cases 
only a part of the whole sky. However, we may construct statistical models with parameters fitted with observations, 
then use the resulting model to generate mock diffuse foreground maps and point source catalogs.  

However, in practice, diffuse maps are usually obtained not at the simulated resolution $\Theta$, but at 
a lower angular resolution $\Theta_0$, as the future experiments being studied are usually designed to have
higher resolution than the existing ones. To address this issue, we shall adopt a simple approach: 
the sky is modeled as a diffuse foreground derived from the fixed angular resolution, plus point sources, i.e. 
\begin{equation}
	I_{\Theta}(\vec{n},\nu)=  I^f_{\Theta_0}(\vec{n},\nu) + \sum_{p \in \textbf{M}(\Theta)} I^p_{\Theta}(\vec{n},\nu) 
	\label{eq:skymodel2}
\end{equation}
For example, for the low frequency sky map we use the Haslam map as a basis, which has $\Theta_0=0.85^{\circ}$. 
The construction of our diffuse foreground model is described in \autoref{sec:diffuse}.

Compared with Eq.\,(\ref{eq:skymodel1}), the model given by  Eq.\,(\ref{eq:skymodel2}) essentially assumes 
that the small scale angular power originates from point sources, whose statistical distribution is assumed to be known, 
and that there is no significant contribution of the extended sources at the scales between $\Theta$ and $\Theta_0$. 
Of course, if the presence of  certain extended sources are known, it may also be possible to generate mock models of these. 
In the present paper we simply ignore such extended sources. Even so, many tests of foreground subtraction techniques or forecast on sky surveys can be conducted using such a model.

The radiation of a source can be detected above the noise if its flux is higher than the sensitivity limit of the observation, 
but whether this radiation can be detected as from a distinct source also depends on the number density of sources.  
Only above a certain threshold, i.e. the confusion limit \cite{2001MNRAS.325.1241V, 2006MNRAS.369..281J, 2012ApJ...758...23C}, 
can the source be distinguished from other sources, below the threshold there are likely multiple faint sources of 
comparable flux within the same beam that they merge into each other 
and become indistinguishable from the diffuse foreground. Thus, in the simulation, we consider point sources
above the confusion limit, while taking the ones below the threshold as part of the diffuse foreground. 

Assuming that the sources are randomly distributed over the sky with a differential source count approximated as a power-law,
\begin{eqnarray}
	n(S) = k\, S^{-\gamma}
	\label{eq:nS}
\end{eqnarray}
For a circular Gaussian beam, the confusion limit flux is \cite{2006MNRAS.369..281J, 2012ApJ...758...23C}
\begin{eqnarray}
	S_c = \left[ \frac{Q^2\, k\, \pi\, \theta_0^2}{(3-\gamma)(\gamma-1)\, 4\ln2} \right]^{\frac{1}{\gamma-1}}
	\label{eq:Sc}
\end{eqnarray}	
where $\theta_0 = 1.029 \lambda/D$, $Q$ is a factor of 3--5 \cite{1973ApJ...183....1M}. 
Note that the confusion limit depends on the angular resolution of the observation and the distribution of sources, but 
not on the sensitivity of the telescope. A detailed computation of the confusion limit is given in \autoref{sec:confusion}. 

As noted above, we may generate the radiation from the set of point sources $\textbf{M}$, which includes all
sources with flux above the confusion limit $S_c$. However, in practice such data may not be available. 
Available surveys may only cover a small part of the sky, and even in the covered regions, the survey may not be complete
up to the confusion limit. However, if a statistical distribution of the point sources can be inferred from the observations, 
we may be able to generate a mock sample of point sources according to the statistical distribution of the sources, 
such that it covers the whole sky up to the confusion limit. 
The aforementioned approach is adopted here: we use the combination of the NVSS and SUMSS surveys. For those regions of sky 
covered by these surveys, especially for the brighter sources which these surveys are essentially complete, 
we will take the point sources from the actual catalog with measured position and flux. At the same time, we also derive the 
statistical distribution of the sources and extrapolate it to faint fluxes. For the regions which are not covered, or for fainter 
sources which the two surveys are incomplete, we will generate mock point sources according to the statistical distribution.

Note that when we constructed the whole sky diffuse maps, except for a few very bright ones, the majority
of the point sources are not individually detected in the input data, so the diffuse map actually included
the radiation from these sources. 
If we are to include the point sources, to avoid double counting, 
we should remove their radiation from the diffuse map to ensure that the model is self-consistent. 
We do this by computing the integrated flux of all the point sources for each pixel of the whole sky map
and subtract them out.  

Below we present the detailed procedures for constructing the diffuse map and the point source contributions.

\begin{figure*}
	\centering  
	\includegraphics[width=1.0\textwidth]{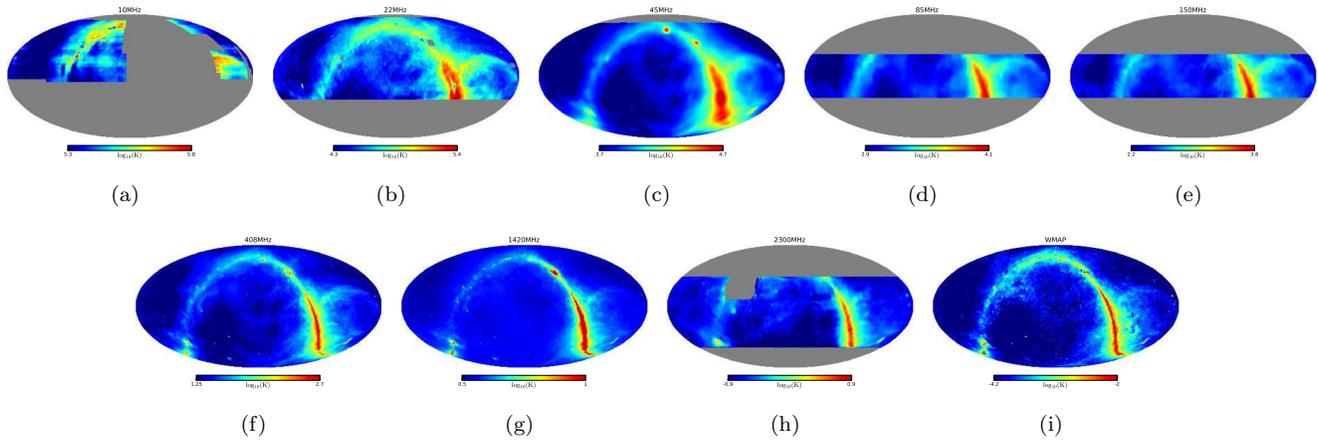}
	\caption{ Sky maps used to model the diffuse foreground. From (a) to (i) are maps at frequencies 
	10 MHz, 22 MHz, 45 MHz, 85 MHz, 150 MHz, 408 MHz, 1420 MHz, 2300 MHz, 23 GHz respectively. }
	\label{fig:diffusedata}
\end{figure*}


\section{Diffuse Emission}
\label{sec:diffuse}

For the whole sky, we pixelize the sky map using the HEALPix scheme \cite{2005ApJ...622..759G}. 
In HEALPix the whole sky is divided into $12 n_{\rm side}^2$ pixels, the pixel size of the map is 
$\theta_{\rm pix} \approx \sqrt{4\pi/(12 n_{\rm side}^2)}$. 
The pixel size should not be larger than the FWHM resolution of the sky map, i.e. $\theta_{\rm pix} \leq \theta_{\rm FWHM}$. 
In the current model we use $n_{\rm side}=512$, which can support resolutions of up to $\sim 0.1^\circ$. 
If higher resolutions are needed, we can pixelize the sky with larger $n_{\rm side}$. 

\begin{table}[H]
	\centering
	\caption{ Sky maps which are used to model the diffuse foreground. }
	\begin{tabular}{|c|c|c|}
	\hline \hline
	Frequency & FWHM & Reference \\
	
	\hline \hline
	10 MHz & $2.6^{\circ} \times 1.9^{\circ}$ & \small \cite{1976MNRAS.177..601C} \\
	\hline
	22 MHz & $1.1^{\circ} \times 1.7^{\circ}$ & \small \cite{1999AandAS..137....7R} \\
	\hline
	45 MHz & $5^{\circ}$ & \small \cite{2011AandA...525A.138G} \\
	\hline
	85 MHz & $3.8^{\circ} \times 3.5^{\circ}$ & \small \cite{1970AuJPA..16....1L} \\
	\hline
	150 MHz & $2.2^{\circ}$ & \small \cite{1970AuJPA..16....1L} \\
	\hline
	408 MHz & $0.85^{\circ}$ & \small \cite{1982AandAS...47....1H} \\
	\hline
	1420 MHz & $0.6^{\circ}$ & \small \cite{2001AandA...376..861R} \\
	\hline
	2300 MHz & $2.3^{\circ} \times 1.9^{\circ}$ & \small \cite{2013yCat..35560001T} \\
	\hline
	\multirow{4}*{\makecell{23, 33,\\ 41, 61,\\ 94 GHz \\ (WMAP)}} & & \\
	& $\sim 0.5^{\circ}$ &\small \cite{2013ApJS..208...20B} \\
	& & \\
	& & \\
	\hline \hline
	\end{tabular}
	\label{tab:diffusedata}
\end{table}

First we construct the whole sky diffuse foreground map. To model the diffuse emission, the GSM uses the observation data of 
whole sky surveys or large area surveys at frequency bands ranging from 
10 MHz to 94 GHz (10 MHz to 5 THz for improved GSM), 
all with different angular resolutions at different frequencies \cite{2008MNRAS.388..247D, 2017MNRAS.464.3486Z}. 
The dataset is listed in \autoref{tab:diffusedata} (only up to 94 GHz for our purpose), \textcolor{red}{and the maps of different 
frequencies are shown in Fig.~\ref{fig:diffusedata}}.
The principal components are obtained by solving the eigenvalue or singular value problem of
the image correlation matrix, and the sky map at arbitrary frequencies are obtained by extrapolating these components, with
the Haslam map at 408 MHz as the base map. As such, the principal components are purely mathematical and do not have
obvious physical origins. 

\begin{figure}[H]
	\centering
	\includegraphics[scale=0.5]{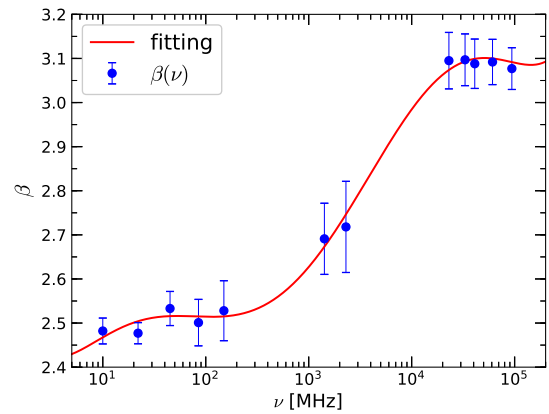}
	\caption{ Polynomial fitting of spectrum index of diffuse emission as frequency.}
	\label{fig:fitfig}
\end{figure}

 However, we consider a simple physical model of 
the foreground radiation, which includes the two major physical components of diffuse foregrounds at low frequencies: 
the galactic synchrotron emission and the galactic free-free emission \cite{1999A&A...345..380S}. 
The galactic synchrotron emission arises from the acceleration of the free high-energy (cosmic ray) electrons in the galactic 
magnetic field. The energy spectrum of the cosmic ray electrons is approximately 
a power-law $E(\gamma) \propto {\gamma}^{-p}$ \cite{2007ARNPS..57..285S, PhysRevD.82.092004, 2011A&A...534A..54S}, 
the brightness temperature of the synchrotron emission is also power-law. 
The galactic free-free (thermal) emission is produced by the scattering of free electrons with diffuse warm ionized gas.
Although it is a small contribution compared with the synchrotron at low frequency \cite{1999A&A...345..380S}, 
its brightness and fluctuation are still much larger than the 21cm signal. 
At high galactic latitude, H$\alpha$ and the diffuse warm ionized gas are optically thin and in local thermodynamic equilibrium, 
they are both proportional to the emission measure (EM), so we can use the galactic H$\alpha$ as a tracer of the
galactic free-free emission \cite{1998astro.ph..1121S}.

The brightness temperature of the diffuse emission can be described as a power law
\begin{eqnarray}
  T_{\rm diffuse}(\nu) = T_{\rm diffuse}(\nu_{*}) \left( \frac{\nu}{\nu_{*}} \right)^{-\beta}
	\label{eq:plf}
\end{eqnarray}
For the spectrum index $\beta$, at the high galactic latitude, \cite{1998ApJ...505..473P} obtained 
$\beta=2.81\pm0.16$ in 1--10 GHz, they then combined it with the Haslam 408 MHz map \cite{1982AandAS...47....1H} and 
Reich \& Reich 1420 MHz map \cite{1988A&AS...74....7R, 2001A&A...368.1123T} to obtain $\beta=2.76\pm0.11$ in 0.4--7.5 GHz. 
At low frequencies, such as 150 MHz, in the same sky region as above, $\beta \approx 2.55$ and increases as 
frequency increases. At about 1 GHz, it steepens to $\beta =$ 2.8--3 \cite{1967MNRAS.136..219B, 1974MNRAS.166..345S, 
1974MNRAS.166..355W, 1979MNRAS.189..465C, 1987MNRAS.225..307L, 1988A&AS...74....7R, 1991MNRAS.248..705B, 1998ApJ...505..473P}.

The spectrum index $\beta$ varies not only with frequency but also with different directions $\vec{n}$.
We have several observations at mid- and low-frequencies from 10 MHz to 2.3 GHz, 
a feasible method to obtain $\beta(\nu,\vec{n})$ is to use the polynomial fitting \cite{2008MNRAS.388..247D}.

\begin{table}[H]
	\centering
	\caption{The fitted $\beta_i$ parameters.}
	\begin{tabular}{cc}
	\hline
	parameter & value and error \\
	\hline
	$\beta_0$  &  $  2.55 \pm 0.035                $ \\
	$\beta_1$  &  $  0.06 \pm 0.002                $ \\
	$\beta_2$  &  $( 3.33 \pm 0.069) \times 10^{-2}$ \\
	$\beta_3$  &  $( 2.57 \pm 0.093) \times 10^{-3}$ \\
	$\beta_4$  &  $( -1.7 \pm 0.028) \times 10^{-3}$ \\
	$\beta_5$  &  $(-9.14 \pm 0.162) \times 10^{-5}$ \\ 
	$\beta_6$  &  $( 2.86 \pm 0.038) \times 10^{-5}$ \\
	\hline
	\end{tabular}
	\label{tab:beta}
\end{table}

Expanding $\beta(\nu,\vec{n})$ in the logarithm of the frequency, 
\begin{eqnarray}
	\ln\left[\frac{T_{\rm diffuse}(\vec{n},\nu)}{T_{\rm diffuse}(\vec{n},\nu_*)}\right] = 
			\beta_0(\vec{n})\left[\ln\frac{\nu}{\nu_*}\right]
	    + \beta_1(\vec{n})\left[\ln\frac{\nu}{\nu_*}\right]^2  +\cdots
	    \nonumber \\
	\label{eq:lgf}
\end{eqnarray}	
rewriting  in matrix form,
\begin{eqnarray}
  y = \textbf{A}\,x + n
	\label{eq:poly}
\end{eqnarray}
For each line of sight, $y$ is a ${\rm N}_{\nu} \times 1$ vector of the data, and
$y=\left[ \ln\frac{T_{\rm diffuse}(\vec{n},\nu_1)}{T_{\rm diffuse}(\vec{n},\nu_*)}, \ln\frac{T_{\rm diffuse}(\vec{n},\nu_2)}{T_{\rm diffuse}(\vec{n},\nu_*)}, \cdots \right]^{\rm T}$, 
$x$ is a ${\rm N}_{\beta} \times 1$ vector of the estimator, and
$x=[\beta_0(\vec{n}), \beta_1(\vec{n}), \cdots]^{\rm T}$, 
\textbf{A} is a ${\rm N}_{\nu} \times {\rm N}_{\beta}$ matrix which is constructed as
\begin{eqnarray}
	\textbf{A} = \begin{bmatrix} \ln\frac{\nu_1}{\nu_*} & \left[\ln\frac{\nu_1}{\nu_*}\right]^2 & \cdots & \left[\ln\frac{\nu_1}{\nu_*}\right]^{{\rm N}_{\beta}} \\
		\ln\frac{\nu_2}{\nu_*} & \left[\ln\frac{\nu_2}{\nu_*}\right]^2 &\cdots & \left[\ln\frac{\nu_2}{\nu_*}\right]^{{\rm N}_{\beta}}\\
			\vdots & & \ddots \\
\ln\frac{\nu_{{\rm N}_{\nu}}}{\nu_*} & \left[\ln\frac{\nu_{{\rm N}_{\nu}}}{\nu_*}\right]^2 & \cdots & \left[\ln\frac{\nu_{{\rm N}_{\nu}}}{\nu_*}\right]^{{\rm N}_{\beta}}
	\end{bmatrix}
	\label{eq:Amatrix}
\end{eqnarray}
where ${\rm N}_{\nu}$ and ${\rm N}_{\beta}$ are the number of frequency bins and the number of parameters required to 
describe the spectrum index of diffuse emission respectively.
The extra term $n$ denotes the noise with $\langle n \rangle=0$.
The estimator $\hat{x}$ is the minimum variance solution of Eq.\,\eqref{eq:poly}
\begin{eqnarray}
  \hat{x} = [\textbf{A}^{\rm T} \textbf{N}^{-1} \textbf{A}]^{-1} \textbf{A}^{\rm T} \textbf{N}^{-1} y
	\label{eq:estimator}
\end{eqnarray}

To compute Eq.\,\eqref{eq:estimator}, we use the observation maps listed in \autoref{tab:diffusedata}, 
the procedure for obtaining the diffuse foreground map is as follows:

\begin{enumerate}
\item Pre-process the observation maps. As the 23 GHz, 33 GHz, 41 GHz, 61 GHz, 94 GHz maps are 
  the WMAP-9yr diffuse radiation map products, we use them directly.  The 2300 MHz map used here has
  already subtracted out the CMB temperature. For the other maps, we subtract the average 
CMB temperature 2.7255 K. The CMB anisotropy \cite{2016A&A...594A...9P} is too small to affect any result. 

\item Subtract the contribution of radio point sources (described in next section) from the maps. 
Note that in the 1420 MHz map some of the strongest sources 
such as Cassiopeia A and Cygnus A have already been removed. 
For each map we generate the point sources as described in \autoref{sec:pointsrc},
first, generate a point source map at the same frequency as the sky map, with a resolution of $45''$; 
second, we use a circular Gaussian function (Eq.\,\ref{eq:circbeam}) or a elliptic Gaussian function (Eq.\,\ref{eq:ellbeam})
to convolve the point source map if the sky map has a circular beam $\theta_0$ or a non-circular beam $\theta_1 \times \theta_2$, 
we then obtained a point source map with the same resolution as the sky map; 
finally we subtract the point source map from the sky map. 

\item Maps produced by following the above two steps (pre-processed maps) represent the diffuse emission. 
In order to compute and fit the spectrum indices, as the input maps are in different angular resolutions, 
all of them are smoothed to a $5^{\circ}$ FWHM resolution, which is the lowest resolution (45 MHz map) among all maps. 
But for the diffuse map simulation, we use the pre-processed 408 MHz map in resolution $0.85^{\circ}$ as the base. 
\label{stepdiffuse}
\end{enumerate}

\begin{figure}[H]
	\centering
	\includegraphics[width=0.44\textwidth]{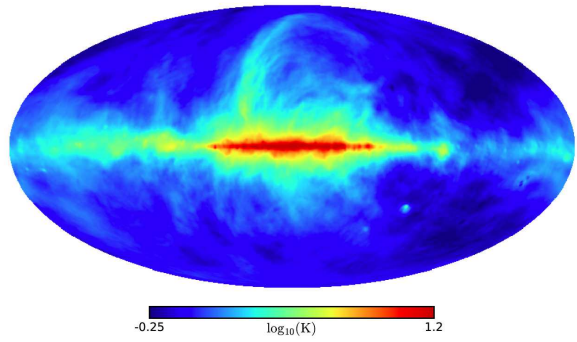}
	\caption{ Simulated diffuse map at 1.4 GHz.}
	\label{fig:diffuse_1400}
\end{figure}

We find that for the expansion in Eq.\,\eqref{eq:lgf},  an order of 6 is a good choice, the fitting curve is shown in 
Fig.\,\ref{fig:fitfig}, and the $\beta_i$\,s are given in \autoref{tab:beta}.

With the spectrum index of diffuse emission fitted above, we can simulate the diffuse emission map 
by scaling the pre-processed 408 MHz map obtained in \autoref{stepdiffuse}. As an example, 
Fig.\,\ref{fig:diffuse_1400} is the simulated diffuse map at 1.4 GHz.


\section{Point Source}
\label{sec:pointsrc}

Most radio point sources are extragalactic. It has been noted that for flux above mJy level, most radio sources are radio-loud 
galaxies (radio-loud AGNs), quasars and BL Lac objects; while at sub-mJy, the star-forming (late-type) galaxies 
(mostly spiral galaxies) dominates \cite{1984ApJ...284...44C, 2005PASA...22...36J}. 
Recent studies suggest that there is another population, radio-quiet AGNs, which are star-forming galaxies but hosting 
an active nucleus \cite{2016ApJ...831..168K, 2017ApJ...842...95M}. 
However, here we are mainly concerned with the flux density distribution???regardless the nature of the sources.

We derive the statistical distribution of the radio point sources from two surveys, 
namely the NRAO VLA Sky Survey (NVSS; \cite{1998AJ....115.1693C}) and the Sydney University Molonglo Sky 
Survey (SUMSS; \cite{1999AJ....117.1578B, 2003MNRAS.342.1117M}). 
The combination of these two surveys covers nearly the entire sky, and their angular resolutions and flux limits are comparable. 
It is more difficult to incorporate other survey data, e.g. the 3C \cite{1959MmRAS..68...37E}, 5C \cite{1975MNRAS.171..475P}, 
FIRST \cite{1995ApJ...450..559B, 2015ApJ...801...26H} surveys, etc., 
because their sky coverages, resolutions and sensitivities are significantly different from each other. 

The NVSS covers the sky of $\delta \ge -40^{\circ}$, which is approximately 82\% of the celestial sphere, 
its observation frequency is 1.4 GHz, with a resolution $\theta_{\rm FWHM} = 45''$ and nearly uniform sensitivity. 
It produces a catalog of almost $2 \times 10^6$ discrete sources stronger than $S \approx 2.5$ mJy. 
SUMSS covers the sky of  $\delta \le -30^{\circ}$, its observation frequency is 843 MHz, and
the angular resolution (FWHM) is $45'' \times 45''{\rm cosec}|\delta|$ where $\delta$ is the declination, 
it produces a catalog of more than $2 \times 10^5$ discrete sources stronger than $S \approx 6$ mJy. 
	
\begin{figure}[H]
	\centering
	\includegraphics[width=0.45\textwidth]{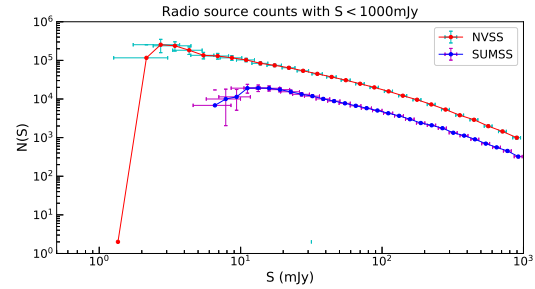}
	\caption{ The source counts of NVSS at 1.4 GHz and SUMSS at 843 MHz. 
	The count drops steeply at $S=2.7$\,mJy for NVSS and $S=12$\,mJy for SUMSS respectively. }
	\label{fig:count}
\end{figure}
\begin{figure}[H]
	\centering
	\includegraphics[width=0.45\textwidth]{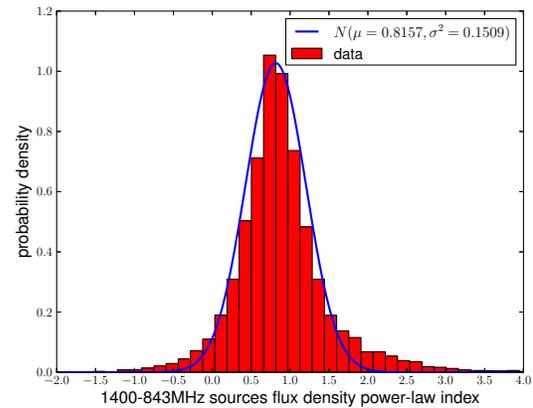}
	\caption{ Distribution of the spectrum index as derived from the sources in the overlap region of NVSS and SUMSS. 
	The distribution is well fit by a Gaussian function $N(\mu=0.8157, \sigma^2=0.1509)$. }
	\label{fig:Sindex}
\end{figure}

In Fig.\,\ref{fig:count} we show the source count distributions $n(S)$ of the two surveys, and we see that
above a certain threshold ($S=2.7$\,mJy for NVSS and $S=12$\,mJy for SUMSS), the $n(S)$ follows a power law distribution. 
The sample of the survey is incomplete when the flux drops below the threshold (sensitivity limit) and 
the density drops steeply at this point. Therefore, in the following analysis we shall only use the sample above the threshold. 

 \begin{figure*}
	\centering
	\includegraphics[width=0.8\textwidth]{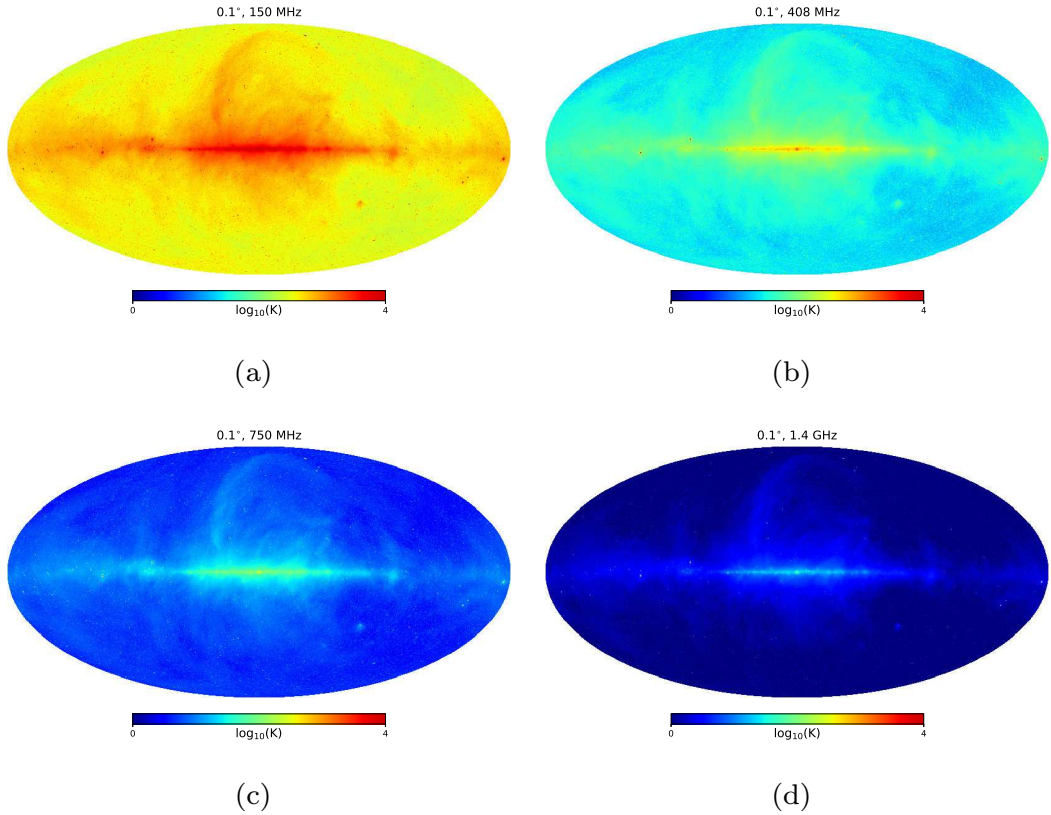}
	\caption{ Maps with resolution of $0.1^{\circ}$ at 150 MHz, 408 MHz, 750 MHz and 1.4 GHz, from (a) to (d). }
	\label{fig:example1d}
\end{figure*}

The survey regions of NVSS and SUMSS overlap for $-40^{\circ} < \delta < -30^{\circ}$,
and the resolutions of NVSS and SUMSS in the overlap region are similar ($\sim 45''$). The sources of the two catalogs 
in the overlap region are cross-identified using the following criterion: if the coordinate difference
is within  $30''$ and there is only one match within the surrounding region of $90''$, the two sources are identified
as the same one. 
In the overlap region,  7515 matched sources are found, in agreement with the SUMSS survey team \cite{2003MNRAS.342.1117M}.
We compute the spectrum index $\alpha$ of these sources from the observations at 1.4 GHz (NVSS) and 843 MHz (SUMSS). 
The distribution of the spectrum index is well fitted by a Gaussian function, 
as shown in Fig.\,\ref{fig:Sindex}, with $\alpha=0.8157 \pm 0.3885$.

With the spectrum index of flux density $\alpha$ fitted above (the index of brightness temperature is $\beta=2+\alpha$), 
we may select a conservative sample limit above which the survey is complete with high confidence. 
For NVSS we set this flux limit as $S = 15$\,mJy, and by scaling the corresponding SUMSS limit is $S = 22$\,mJy, 
i.e. $S_{1.4\,{\rm GHz}} = S_{843{\rm MHz}} \left(\frac{1.4\,{\rm GHz}}{843\,{\rm MHz}}\right)^{-\alpha}$ where $\alpha=0.8157$, 
the surface densities of sources in the two samples are then similar. We assume the NVSS and SUMSS surveys obtained
complete samples of low frequency radio sources above this limit, then by combining the two surveys, 
we obtain a complete catalog of radio point sources above this limit for most of the sky. Thus, if the point source catalog
is to be limited above 15 mJy at 1.4 GHz, we can use this actual catalogue to generate the point sources.

However, the confusion limit for our input all-sky map is well below this value, 
which is only 0.5 mJy (see \autoref{sec:confusion}). 
We need to take into account of the point sources in the flux range of 0.5--15 mJy. For these sources
the complete catalog is not available, but we can generate mock point sources according to their statistical distribution.
The $n(S)$ distribution can be derived from the existing data, as detailed in \autoref{sec:ns}. 
For a sky region whose solid angle is $\Omega$, the total number of sources is $N_{\rm tot}=\int_S n(S) \Omega\, {\rm d}S$. 
There are about $2.5 \times 10^8$ such sources in the whole sky, its distribution can be written approximately as 
\begin{eqnarray}
	n(S) \approx 1300\, S^{-1.77}
\end{eqnarray}

The simplest way for generating the mock point sources would be to place them with a uniform random distribution 
on the celestial sphere. However, it has been observed that there is angular clustering among the radio sources, as would be 
expected for general matter distribution \cite{2003A&A...405...53O, 2004MNRAS.355.1053D}. To generate point sources with 
such angular clustering, we use the Rayleigh-L\'evy random walk method, in which the distribution of the angular distance 
is \cite{1975CRASM.280.1551M, 1995ApJS...96..401C, 2003AJ....125.2064G, 2008MNRAS.389.1319J, 2014MNRAS.440...10A}
\begin{eqnarray}
	P(>\theta) = \left\{ \begin{array}{ll}
	\left(\frac{\theta}{\theta_0}\right)^{-\gamma} \,, & \quad \theta \geq \theta_0 \\ 
	\quad 1 \,, & \quad \theta < \theta_0
	\end{array} \right.
	\label{eq:flight}
\end{eqnarray}
The parameters $\theta_0$ and $\gamma$ can be obtained from the two-point correlation function between NVSS and FIRST. 
Here we follow the result from \cite{2003A&A...405...53O} to use $\theta_0=6'$ and $\gamma=0.8$.

We execute the following steps to generate the point sources of the Rayleigh-L\'evy random walk:  
\begin{enumerate}
\item Start from an arbitrary position, place the first source;

\item Choose a random direction $\vec{n}$, a random angular distance $\theta$ and an uniform random number $u$. 
If $u > P(>\theta)$, then place the next source at the coordinate $(\vec{n},\theta)$;

\item Repeat the Step 2, until all sources are placed; 

\item Specify the flux of each source at 1.4 GHz. 
For each flux bin centered at $S_0$ with width $\Delta S_0$, the number of sources over the whole sky is 
$N_0 = 4\pi \int_{S_0-\Delta S_0/2}^{S_0+\Delta S_0/2} n(S)\,{\rm d}S$. 
Therefore, we randomly select $N_0$ sources from the mock map, 
give flux densities as uniform random values from $S_0-\Delta S_0/2$ to $S_0+\Delta S_0/2$, 
with the spectrum index given by a Gaussian random number with $N(\mu=0.8157, \sigma^2=0.1509)$.
\end{enumerate}	
	
Indeed, even for flux density above 15 mJy at 1.4 GHz, there are a few gaps on the sky not covered by the combined NVSS+SUMSS catalog. The same procedure can be used to generate sources which fill these gaps.


\section{Results and Discussions}
\label{sec:result}

Now we can make the total sky map by summing up the diffuse foreground map (which is itself the sum of Galactic synchrotron 
and galactic free-free emission map) and the point sources. 
In the sky regions covered by the NVSS and SUMSS surveys, the bright sources
are from the actual catalog. For sky regions not covered by the surveys, or for the fainter sources, point sources are generated 
according to statistical distribution. Using the spectrum indices we obtained for the diffuse maps and the point sources,  
the flux at the desirable frequencies are computed, and then convolved with a beam of assumed resolution. 

\textcolor{red}{We produce the total sky maps at 150 MHz, 408 MHz, 750 MHz, 1.4 GHz, 
and 2.3 GHz with FWHM resolutions $0.1^{\circ}$ and $0.015^{\circ}$ (shown in Fig.~\ref{fig:example1d
}), as well as a point source catalogue above 0.5 mJy at 1.4 GHz. The sky appears brighter at lower frequencies, but 
in the frequency range discussed here the large features are similar.}
Our sky maps and point source catalogs are available for download\footnote{\url{http://tianlai.bao.ac.cn/~huangqizhi/SSM/}}.

In the sky regions covered by the NVSS and SUMSS surveys, the brighter sources
are from the actual catalogue. For sky regions not covered by the surveys, or for the fainter sources, point sources are generated 
according to statistical distribution. Using the spectrum indices we obtained for the diffuse maps and the point sources,  
the flux at the desirable frequencies are computed, and then convolved with a beam of assumed resolution. 

\begin{figure}[H]
	\centering
	\includegraphics[width=0.4\textwidth]{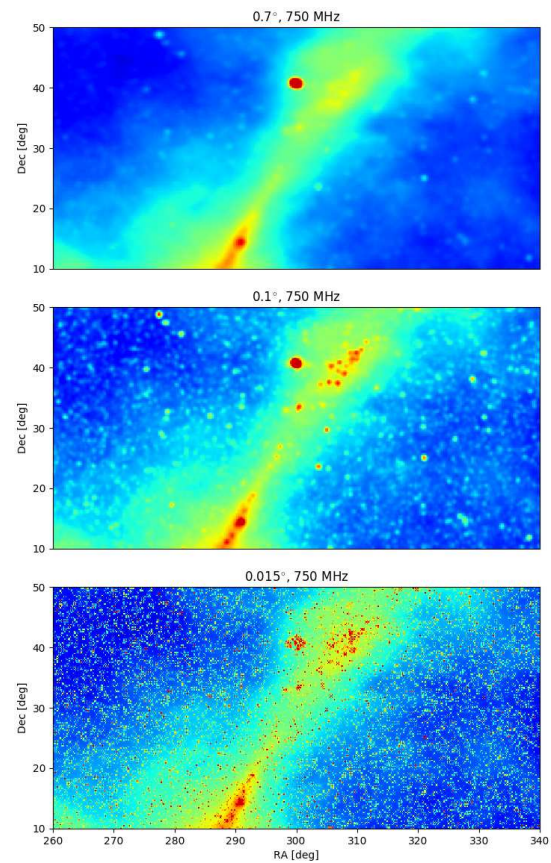}
	\caption{ Zoomed-in sky maps in a region of RA from $260^{\circ}$ to $340^{\circ}$ 
	and Dec from $10^{\circ}$ to $50^{\circ}$. }
	\label{fig:example750zoom}
\end{figure}

In Fig.\,\ref{fig:example750zoom} we show a zoom up at the mid-declination for the maps at the different 
resolutions of $0.7^\circ$, $0.1^\circ$, and $0.015^\circ$. We can see that as the resolution improved, 
more radio sources can be seen clearly in the map, some sources which are merged in the low resolution
maps appear distinct at higher resolutions, while the overall temperature of the map remains the same.

\begin{figure}[H]
	\centering
	\includegraphics[width=0.4\textwidth]{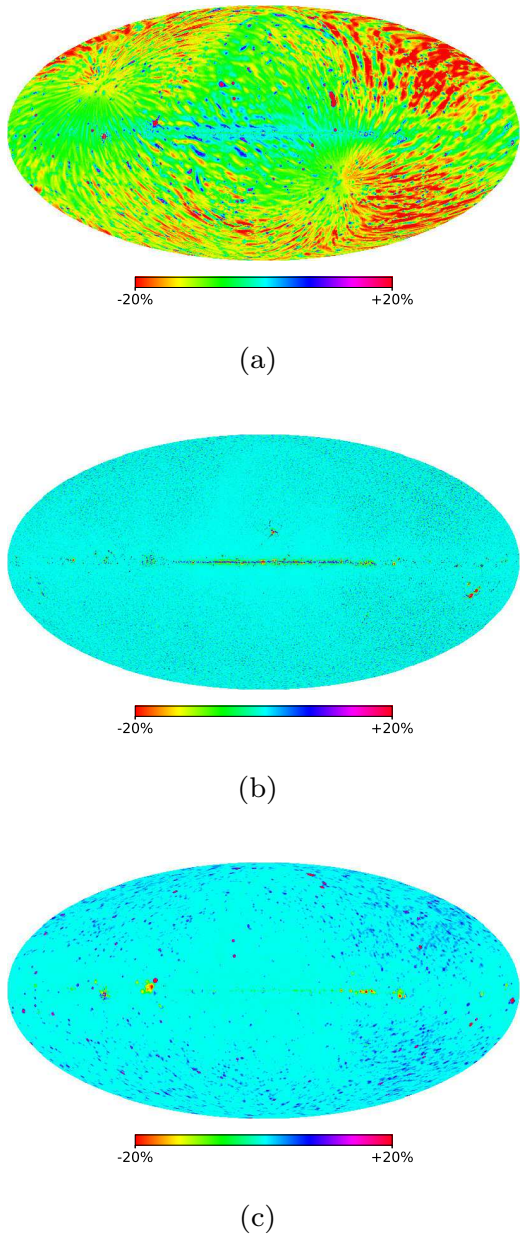}
	\caption{ Relative differences of the GSM2008 (a), GSM2016 (b),  and our model (c)
	to the 2014 version of Haslam map.}
	\label{fig:err}
\end{figure}

In Fig.\,\ref{fig:err} we plot the relative differences of the GSM (2008 and 2016) and our model with respect
to the 2014 version of the Haslam map at 408 MHz in a resolution of $\sim$$1^{\circ}$.
Compared with the original 1982 version, in 
the 2014 version of the Haslam map \cite{2015MNRAS.451.4311R} strong baseline stripings are removed, however
the stronger radio sources are also removed. So we can see that the GSM 2008 model which is based on the original 1982
version of Haslam map show differences in the form of stripes and points. The GSM 2016 model is itself based on
the 2014 version of Haslam map, so there is very small difference. Our own model also has
small differences with respect to the 2014 version of 
Haslam map, but as we added back the point sources, we can see some differences appearing as points.

\begin{figure}[H]
	\centering 
	\includegraphics[width=0.47\textwidth]{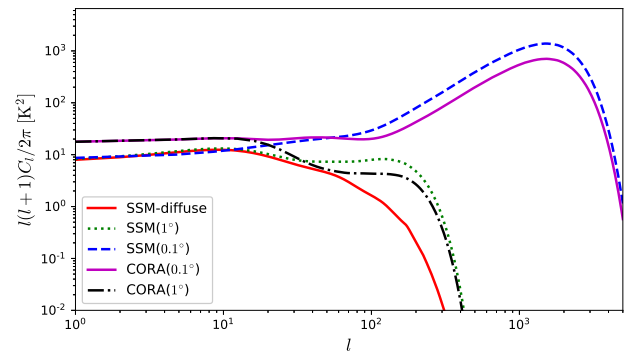}
	\caption{ The angular power spectra of the diffuse and total power maps at $1^\circ$ and $0.1^\circ$ resolutions. 
	For comparison, we also plot the angular power spectrum  from a map generated by the CORA code. } 
	\label{fig:cl}
\end{figure}

With the addition of the point sources, we expect significant changes in the angular power spectrum. In Fig.\,\ref{fig:cl} 
we plot the angular power spectrum for both our diffuse map and total map, at $1^\circ$ and $0.1^\circ$ resolutions. 
For comparison, we also use the CORA code to generate mock maps at these resolutions and plot the resulting angular 
power spectrum in this figure. 
For the diffuse map, the angular power is basically flat until it begins to drop at the scale corresponding to the map resolution. 
With the addition of the point sources, there are significant powers on the small scales. The angular power spectrum actually 
raises before reaching the map resolution. 
The general shape of our angular power spectrum is similar to that of the CORA map, 
though there are slight differences in the amplitude. As the two maps are both generated with random distribution, and there are
differences in the details of both the assumed distribution and the method of generation, such difference  
is not unexpected. 
Compared with the low resolution GSM model,  such angular power will have substantial impacts on the foreground
removal results. 

Note however our model consists a low resolution diffuse sky model and
a point source catalog model, but no high resolution extended sources, so the angular power spectrum may not 
be completely accurate. Extended sources do exist, e.g. some galactic 
supernova remnants. Many extragalactic sources, e.g. radio galaxies, radio halo of galaxy clusters and groups, etc., 
are also extended source under higher resolutions. Such extended sources could be modeled in a manner similar to our current 
modeling of point sources but with more parameters each: in addition to the center location coordinates, 
total flux and spectral index, also the parameters describing its 
shapes (modeled e.g. with one or more elliptic Gaussian profiles) \cite{2018arXiv180505222B}.
 
There are some limitations in the model presented here. In terms of applicable frequency range, here 
we considered only below 2.3 GHz. At this mid-to-low frequency range, the main components of the 
foreground radiation can be modeled using the galactic synchrotron emission, galactic free-free
emission and the extragalactic sources 
which are also primarily radiating by these mechanism. We have not considered 
the propagation effects (absorption, refraction and scattering) in this treatment. At the very low frequencies these 
effects may have significant impacts. For example, the radiowaves at a few MHz may suffer significant absorption because of the
interstellar medium electrons. These shall be studied in the future.

We have considered only unpolarized emission in this study. In fact, the galactic synchrotron emission is highly polarized,
and some  
point sources are also polarized \cite{2018A&A...613A..58V}. The polarized emission is also very important in the foreground 
subtraction analysis, because the Faraday rotation induce an oscillation along the frequency which is similar to the 21cm signal
\cite{2010MNRAS.409.1647J, Asad:2015sda, 2016MNRAS.462.4482A, 2017A&A...597A..98V, 2018MNRAS.476.3051A, Spinelli:2018kpm}. 

Despite these limitations, the inclusion of the point source in this model increased the usable range of angular resolution, 
making it a useful tool for the study of foreground subtraction in 21cm experiment. Our model can be further improved in the 
future by implementing  the extended sources, polarization and propagation effects.


{\it
We thank R\'eza Ansari for discussions. 
The computations of this paper is performed on the Tianhe-2 supercomputer at the Guangzhou supercomputing center with the 
support of NSFC-Guangdong Joint grant U1501501. This research is supported by the Ministry of Science and Technology grant 
2016YFE0100300, NSFC grants 11473044, 11761141012, 11633004, 11653003, and CAS grant QYZDJ-SSW-SLH017. 
F. Q. Wu also acknowledge the support by the CSC Cai Yuanpei grant. }

\InterestConflict{The authors declare that they have no conflict of interest.}


\bibliographystyle{abbrv,sort}

\begin{appendix}

\section{Computation of the Confusion Limit}
\label{sec:confusion}

The beam of a telescope can be modeled as a circular Gaussian function 
\begin{eqnarray}
	B_{\rm circ}(\theta,\varphi) = \exp \left\{ -4 \ln2\, \frac{\theta^2}{\theta_0^2} \right\} \, , \quad
	\theta_0 = 1.029 \frac{\lambda}{D}
	\label{eq:circbeam}
\end{eqnarray}
where $\theta_0$ is the FWHM resolution, $\lambda$ is the observation wavelength,
$D$ is the aperture size of the telescope, 
or an elliptic Gaussian function 
\begin{eqnarray}
	B_{\rm ell}(\theta, \varphi) = \exp\left\{ -4\ln2\, \theta^2 \left[ \left(\frac{\cos\varphi}{\theta_1}\right)^2 + 
			\left(\frac{\sin\varphi}{\theta_2}\right)^2 \right] \right\}
	\label{eq:ellbeam}
\end{eqnarray}

For a telescope with a beam response $B(\theta,\varphi)$, a source with flux $S$ will induce an output  
$x=S \cdot B(\theta,\varphi)$, the average number of response $R(x)$ with an amplitude
between $x$ and $x+{\rm d}x$ in a solid angle $\Omega_b$ is then
\cite{2001MNRAS.325.1241V, 2006MNRAS.369..281J, 2012ApJ...758...23C}
\begin{eqnarray}
	R(x) = \int_{\Omega_b} \frac{n\left( \frac{x}{B(\theta,\varphi)} \right)}{B(\theta,\varphi)} {\rm d}\Omega
	\label{eq:Rx}
\end{eqnarray}
For a circular Gaussian beam, the integration of Eq.\,\eqref{eq:Rx} is given by
\begin{eqnarray}
	R(x) = k\, x^{-\gamma}\, \frac{\pi\, \theta_0^2}{(\gamma-1)\,4 \ln2}
	\label{eq:Rx2}
\end{eqnarray}
Let $S_c=Q\sigma_c$ where Q is a factor of 3--5 \cite{1973ApJ...183....1M}, the rms confusion $\sigma_c$ is 
\cite{2006MNRAS.369..281J, 2012ApJ...758...23C}
\begin{eqnarray}
	\sigma_c^2 = \int_0^{S_c} x^2 R(x) {\rm d}x
	\label{eq:confusionintf}
\end{eqnarray}
If the differential count of radio sources $n(S)$ is a single power-law as given in Eq.\,(\ref{eq:nS}), 
Eq.\,\eqref{eq:confusionintf} becomes
\begin{eqnarray}
	S_c = \left[ \frac{Q^2\, k\, \pi\, \theta_0^2}{(3-\gamma)(\gamma-1)\, 4\ln2} \right]^{\frac{1}{\gamma-1}}
\end{eqnarray}	
For broader range of frequencies, we may model  $n(S)$ using a piece-wise power law \cite{1998AJ....115.1693C, 2003MNRAS.342.1117M}, 
\begin{eqnarray}
	S^{5/2} n(S) = k_n\, S^{5/2 - \gamma_{_n}}, \quad n=1,2,3,4,5,\cdots
	\label{eq:multif}
\end{eqnarray}
The best-fit parameters are 

\begin{eqnarray}\begin{array}{lrl}
	\gamma_1=1.57, \quad k_1=29500 & \quad {\rm for} & \quad\qquad\quad\,\, S \leq 10^{-4} \,\,{\rm Jy} \\
	\gamma_2=2.23, \quad k_2=60    & \quad {\rm for} & \quad       10^{-4}< S \leq 10^{-3} \,\,{\rm Jy} \\
	\gamma_3=1.77, \quad k_3=1300  & \quad {\rm for} & \quad       10^{-3}< S \leq 10^{-1} \,\,{\rm Jy} \\
	\gamma_4=2.20, \quad k_4=440   & \quad {\rm for} & \quad       10^{-1}< S \leq 1 \,\,\,\,\quad{\rm Jy} \\
	\gamma_5=2.77, \quad k_5=390   & \quad {\rm for} & \quad\qquad\quad\,\, S    > 1 \,\,\,\,\quad{\rm Jy}
	\end{array}
	\label{eq:kgamma}
\end{eqnarray}

With these parameters we obtain the confusion limit, we find
\begin{eqnarray}
	S_c = \left\{ \begin{array}{rl}
		0.086 \,{\rm mJy} \cdot Q^{3.51} \left(\frac{\theta_0}{\rm arcmin}\right)^{3.51} \left(\frac{\nu}{{\rm GHz}}\right)^{-2\alpha} \,, & \quad\qquad\quad\,\, \theta_0 \leq 0.25' \nonumber\\
		 0.12 \,{\rm mJy} \cdot Q^{1.53} \left(\frac{\theta_0}{\rm arcmin}\right)^{1.53} \left(\frac{\nu}{{\rm GHz}}\right)^{-2\alpha} \,, & \quad 0.25' < \theta_0 \leq 1.13' \nonumber \\
		0.017 \,{\rm mJy} \cdot Q^{2.56} \left(\frac{\theta_0}{\rm arcmin}\right)^{2.56} \left(\frac{\nu}{{\rm GHz}}\right)^{-2\alpha} \,, & \quad1.13' < \theta_0 \leq 7.18'  \nonumber\\
		 0.51 \,{\rm mJy} \cdot Q^{1.63} \left(\frac{\theta_0}{\rm arcmin}\right)^{1.63} \left(\frac{\nu}{{\rm GHz}}\right)^{-2\alpha} \,, & \quad 7.18' < \theta_0 \leq 30' \nonumber\\
		    5 \,{\rm mJy} \cdot Q^{1.18} \left(\frac{\theta_0}{\rm arcmin}\right)^{1.18} \left(\frac{\nu}{{\rm GHz}}\right)^{-2\alpha} \,, & \quad\qquad\quad\,\, \theta_0 > 30' \nonumber
\end{array} \right.
\label{eq:confusionlimit}
\end{eqnarray}
where $\alpha$ is the spectrum index of flux for the sources, typically $\alpha=$ 0.7--0.8 (Fig.\,\ref{fig:Sindex}), 
For example, for the NVSS+SUMSS catalogue, $\theta_0=45'' =0.75'$, if choose $Q=5$, 
the confusion limit $S_c=0.5$ mJy and the rms $\sigma_c = 0.1$ mJy. So
point source whose flux density above 0.5 mJy at 1.4 GHz can be distinguished.


\section{Estimation of Differential Surface Density of Sources}
\label{sec:ns}

In order to calculate $n(S)$, we first compute the radio luminosity function. 
Two methods are usually used to calculate the local luminosity function, the maximum volume $V_m$ method 
\cite{1984ApJ...284...44C, 1987A&A...184....7T} and the STY maximum-likelihood method
\cite{1979ApJ...232..352S}. Here we use the maximum volume $V_m$ method, 
in which the target source is moved through the redshifts to find the maximum volume it could have inhabited and
still been observed: 
\begin{eqnarray}
	\rho_m(L) = \sum_{i=1}^N \left( \frac{1}{V_m} \right)_i
	\label{eq:rho_m}
\end{eqnarray}
where $N$ is the number of sources. Assuming the distribution of sources is random, the root mean square (r.m.s.) error is
\begin{eqnarray}
	\sigma = \sqrt{ \sum_{i=1}^N \left( \frac{1}{V_m} \right)_i^2 }
	\label{eq:rho_m_err}
\end{eqnarray}
We use the data from \cite{2007MNRAS.375..931M}, with the cosmological parameters from \cite{2016A&A...594A..13P}. 

For start-forming galaxies, the luminosity function is of the form\cite{1990MNRAS.242..318S}
\begin{eqnarray}
	\rho_m(L) = C \left[ \frac{L}{L_*{}} \right]^{1-a} \exp\left\{ -\frac{1}{2} \left[ \frac{\log_{10}\left( 1+\frac{L}{L_{*}} \right)}{\sigma} \right]^2 \right\}
	\label{eq:SFf}
\end{eqnarray}
The best-fit parameters of Eq.\,\eqref{eq:SFf} are
\begin{eqnarray}
	C      & = & 10^{-2.88 \pm 0.74}\, {\rm mag}^{-1} {\rm Mpc}^{-3}  \nonumber \\
	L_{*}  & = & 10^{21.22 \pm 5.67}\,\, {\rm W\, Hz}^{-1}              \nonumber \\
	a      & = & 1.04 \pm 0.21                                                  \\
	\sigma & = & 0.61 \pm 0.14                                        \nonumber
	\label{eq:sffit}
\end{eqnarray}

For radio-loud AGNs, the luminosity function is of the form\cite{1989ApJ...338...13C}
\begin{eqnarray}
	\log_{10}(\rho_m) = Y - \frac{3}{2}\log_{10}(L) - \sqrt{ B^2 + \left[ \frac{\log_{10}(L) - X}{W} \right]^2 }
	\label{eq:AGNf}
\end{eqnarray}
The best-fit parameters of Eq.\,\eqref{eq:AGNf} are
\begin{eqnarray}
	Y & = & 33.89 \pm 7.34  \nonumber \\
	B & = &  2.27 \pm 0.83  \nonumber \\
	X & = & 25.96 \pm 2.27            \\
	W & = &  0.85 \pm 0.12  \nonumber 
	\label{eq:agnfit}
\end{eqnarray}

\begin{figure}[htbp]
	\centering
	\includegraphics[width=0.46\textwidth]{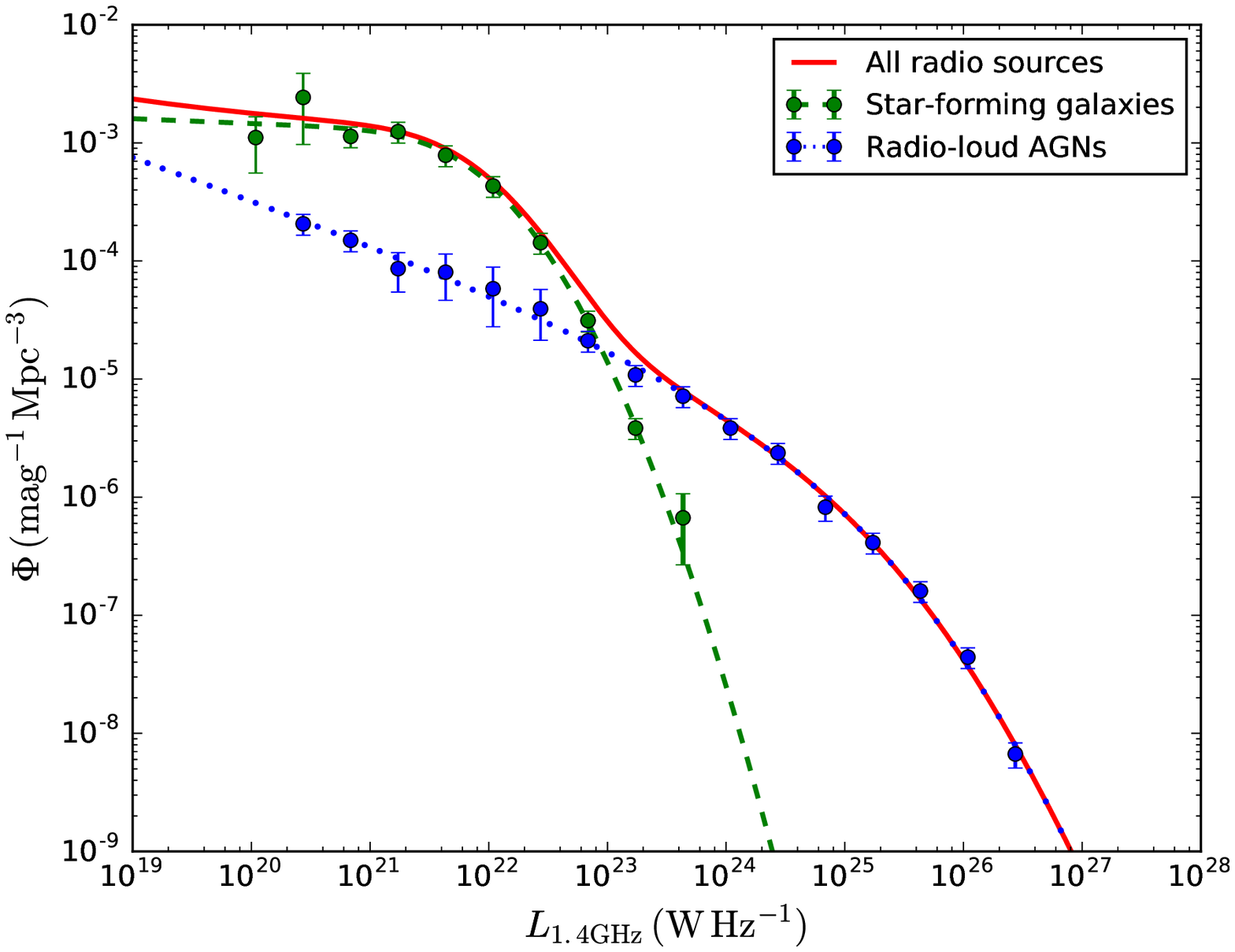}
	\caption{ The radio local luminosity function at 1.4 GHz. 
	Luminosity below $10^{22}\, {\rm W\, Hz^{-1}}$, star-forming galaxies are dominated (green dash line), 
	but above $10^{24}\, {\rm W\, Hz^{-1}}$, radio-loud AGNs become dominated (blue dot line). }
	\label{fig:RLF}
\end{figure}

\autoref{fig:RLF} shows the fitted local radio luminosity function.

For an isotropic source with spectrum index $\alpha \equiv-{\rm d}[\ln(S)]/{\rm d}[\ln(\nu)]$ at redshift $z$, its flux is given by 
\begin{eqnarray}
	S = \frac{L}{4\pi D_L^2 (1+z)^{1+\alpha}} = \frac{L}{A (1+z)^{1+\alpha}}
	\label{eq:S-L}
\end{eqnarray}
where $L$ is the luminosity, and the luminosity distance of this source is
$D_L = \frac{c(1+z)}{H_0} \int_0^z \frac{{\rm d}z^{'}}{\sqrt{\Omega_{\Lambda} + \Omega_m(1+z^{'})^3}}$,
and we adopt the $\Lambda$CDM cosmological model with the parameters from Plank 2015 \cite{2016A&A...594A..13P}.
Then we have
\begin{eqnarray}
	\frac{{\rm d}L}{{\rm d}S} = A(1+z)^{1+\alpha}
	\label{eq:dLdS}
\end{eqnarray}

In a comoving distance $r$ to $r+{\rm d}r$, the number of sources with luminosities $L$ to $L+{\rm d}L$ is 
$\rho_m(L,z)A{\rm d}L{\rm d}r$. 
This number can also be expressed by the sources with flux densities $S$ to $S+{\rm d}S$ in redshift range $z$ to $z+{\rm d}z$,  
that is $\eta(S,z){\rm d}S{\rm d}z$. The comoving distance is given by $d r = c dz/H(z)$.
Let $\eta(S,z){\rm d}S{\rm d}z = \rho_m(L,z)A{\rm d}L{\rm d}r$, we obtain
\begin{eqnarray}
	\eta(S,z) = \rho(L,z) A \frac{{\rm d}L}{{\rm d}S} \frac{{\rm d}r}{{\rm d}z} = \frac{cA^2 (1+z)^{1+\alpha} \rho(L,z)}{H_0 \sqrt{\Omega_{\Lambda} + \Omega_m(1+z)^3}}
	\label{eq:etaf}
\end{eqnarray}
	
The weighted differential count is 
\begin{eqnarray}
	S^{5/2} n(S) = \frac{1}{4\pi} \int_0^{\infty} S^{5/2} \eta(S,z) {\rm d}z
	\label{eq:differentcountf}
\end{eqnarray}
where $n(S)$ is the number of sources per steradian per flux density.
	
\begin{figure}[H]
	\centering
	\includegraphics[width=0.49\textwidth]{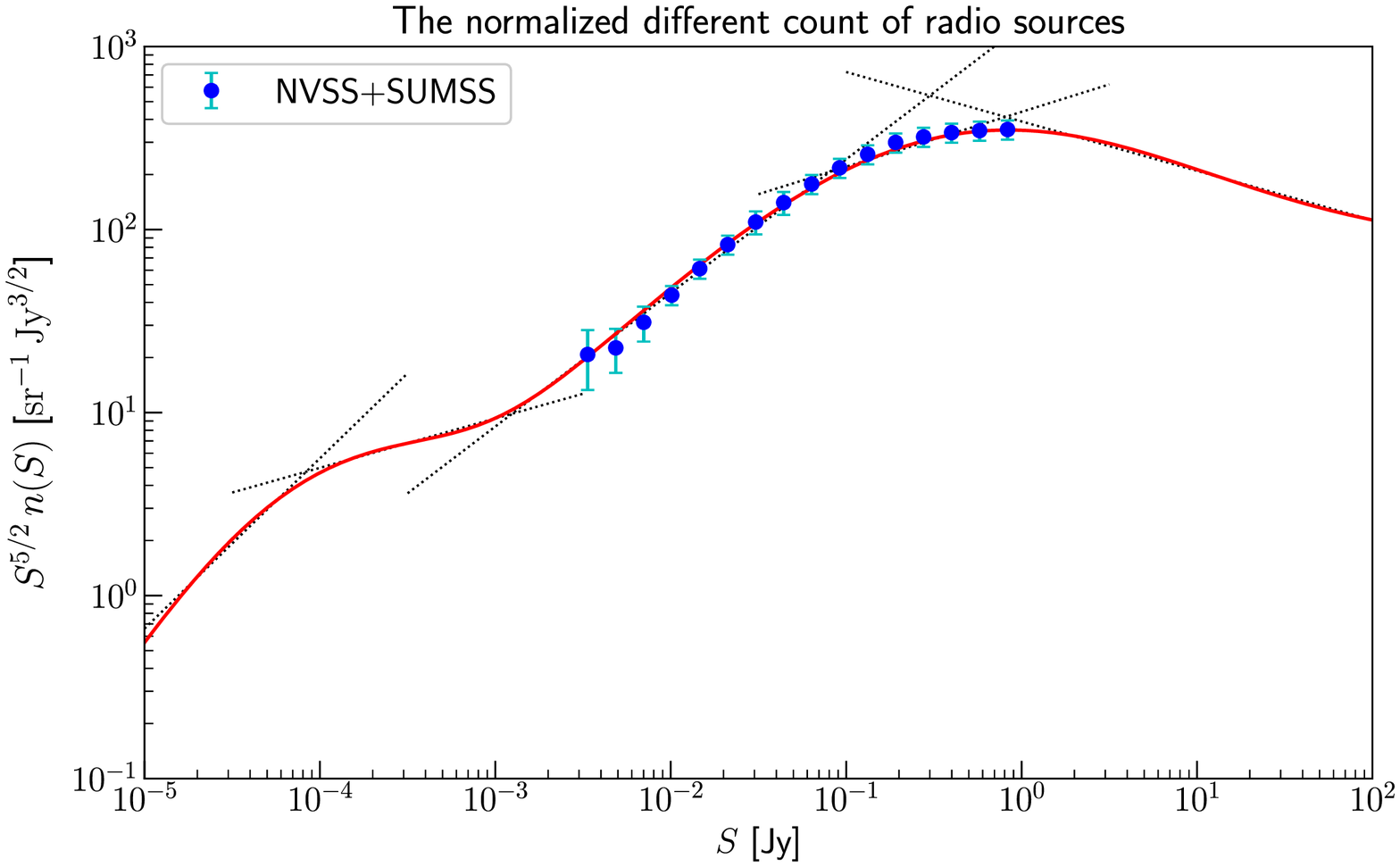}
	\caption{ The weighted different count $S^{5/2} n(S)$ of radio sources. 
	The blue points are calculated from the NVSS-SUMSS full-sky complete sample, 
	and the red curve is a plot of Eq.\,\eqref{eq:differentcountf}. }
	\label{fig:normdiffcount}
\end{figure}	
	
We model the redshift evolution of the radio luminosity function  as a combination of 
``pure luminosity evolution'' and ``pure density evolution'' \cite{1984ApJ...284...44C},
\begin{eqnarray}
	\rho_m(L,z) = g(z)\, \rho_m \left( \frac{L}{f(z)},0 \right)
	\label{eq:Lz}
\end{eqnarray}
where
\begin{eqnarray}
	f(z) = (1+z)^x \,, \quad g(z) = (1+z)^y \, e^{-\left(\frac{z}{z_c}\right)^q}
	\label{eq:gzcut}
\end{eqnarray}	
We use the full-sky complete sample from NVSS and SUMSS to fit the evolution of radio luminosity function, 
the weighted different count of radio sources (Eq.\,\ref{eq:differentcountf}) is shown in Fig.\,\ref{fig:normdiffcount}, and
the best-fit parameters are
\begin{eqnarray}
 	  x & = &  6.68 \pm 1.33  \nonumber \\
	  y & = & -8.34 \pm 1.86  \nonumber \\
	z_c & = &  2.39 \pm 0.17            \\
	  q & = &  2.53 \pm 0.24  \nonumber
\end{eqnarray}
If approximate as a single power law, we dind $n(S) \approx 1300\, S^{-1.77}$.

\end{appendix}


\begin{thebibliography}{10}

\bibitem{1997ApJ...475..429M}
P.~{Madau}, A.~{Meiksin}, and M.~J. {Rees}.
\newblock {21 Centimeter Tomography of the Intergalactic Medium at High
  Redshift}.
\newblock {\em ApJ}, 475:429--444, February 1997.

\bibitem{2003ApJ...596....1C}
B.~{Ciardi} and P.~{Madau}.
\newblock {Probing beyond the Epoch of Hydrogen Reionization with 21 Centimeter
  Radiation}.
\newblock {\em ApJ}, 596:1--8, October 2003.

\bibitem{2015MNRAS.447.1806G}
R.~{Ghara}, T.~R. {Choudhury}, and K.~K. {Datta}.
\newblock {21 cm signal from cosmic dawn: imprints of spin temperature
  fluctuations and peculiar velocities}.
\newblock {\em MNRAS}, 447:1806--1825, February 2015.

\bibitem{2008PhRvL.100i1303C}
T.-C. {Chang}, U.-L. {Pen}, J.~B. {Peterson}, and P.~{McDonald}.
\newblock {Baryon Acoustic Oscillation Intensity Mapping of Dark Energy}.
\newblock {\em Physical Review Letters}, 100(9):091303, March 2008.

\bibitem{2008PhRvD..78b3529M}
Y.~{Mao}, M.~{Tegmark}, M.~{McQuinn}, M.~{Zaldarriaga}, and O.~{Zahn}.
\newblock {How accurately can 21cm tomography constrain cosmology?}
\newblock {\em Phys. Rev.~D}, 78(2):023529, July 2008.

\bibitem{2004PhRvL..92u1301L}
A.~{Loeb} and M.~{Zaldarriaga}.
\newblock {Measuring the Small-Scale Power Spectrum of Cosmic Density
  Fluctuations through 21cm Tomography Prior to the Epoch of Structure
  Formation}.
\newblock {\em Physical Review Letters}, 92(21):211301, May 2004.

\bibitem{2013A&A...556A...2V}
M.~P. {van Haarlem}, M.~W. {Wise}, A.~W. {Gunst}, G.~{Heald}, J.~P. {McKean},
  J.~W.~T. {Hessels}, A.~G. {de Bruyn}, R.~{Nijboer}, J.~{Swinbank},
  R.~{Fallows}, M.~{Brentjens}, A.~{Nelles}, R.~{Beck}, H.~{Falcke},
  R.~{Fender}, J.~{H{\"o}randel}, L.~V.~E. {Koopmans}, G.~{Mann}, G.~{Miley},
  H.~{R{\"o}ttgering}, B.~W. {Stappers}, R.~A.~M.~J. {Wijers}, S.~{Zaroubi},
  M.~{van den Akker}, A.~{Alexov}, J.~{Anderson}, K.~{Anderson}, A.~{van
  Ardenne}, M.~{Arts}, A.~{Asgekar}, I.~M. {Avruch}, F.~{Batejat},
  L.~{B{\"a}hren}, M.~E. {Bell}, M.~R. {Bell}, I.~{van Bemmel}, P.~{Bennema},
  M.~J. {Bentum}, G.~{Bernardi}, P.~{Best}, L.~{B{\^i}rzan}, A.~{Bonafede},
  A.-J. {Boonstra}, R.~{Braun}, J.~{Bregman}, F.~{Breitling}, R.~H. {van de
  Brink}, J.~{Broderick}, P.~C. {Broekema}, W.~N. {Brouw}, M.~{Br{\"u}ggen},
  H.~R. {Butcher}, W.~{van Cappellen}, B.~{Ciardi}, T.~{Coenen}, J.~{Conway},
  A.~{Coolen}, A.~{Corstanje}, S.~{Damstra}, O.~{Davies}, A.~T. {Deller}, R.-J.
  {Dettmar}, G.~{van Diepen}, K.~{Dijkstra}, P.~{Donker}, A.~{Doorduin},
  J.~{Dromer}, M.~{Drost}, A.~{van Duin}, J.~{Eisl{\"o}ffel}, J.~{van Enst},
  C.~{Ferrari}, W.~{Frieswijk}, H.~{Gankema}, M.~A. {Garrett}, F.~{de
  Gasperin}, M.~{Gerbers}, E.~{de Geus}, J.-M. {Grie{\ss}meier}, T.~{Grit},
  P.~{Gruppen}, J.~P. {Hamaker}, T.~{Hassall}, M.~{Hoeft}, H.~A. {Holties},
  A.~{Horneffer}, A.~{van der Horst}, A.~{van Houwelingen}, A.~{Huijgen},
  M.~{Iacobelli}, H.~{Intema}, N.~{Jackson}, V.~{Jelic}, A.~{de Jong},
  E.~{Juette}, D.~{Kant}, A.~{Karastergiou}, A.~{Koers}, H.~{Kollen}, V.~I.
  {Kondratiev}, E.~{Kooistra}, Y.~{Koopman}, A.~{Koster}, M.~{Kuniyoshi},
  M.~{Kramer}, G.~{Kuper}, P.~{Lambropoulos}, C.~{Law}, J.~{van Leeuwen},
  J.~{Lemaitre}, M.~{Loose}, P.~{Maat}, G.~{Macario}, S.~{Markoff},
  J.~{Masters}, R.~A. {McFadden}, D.~{McKay-Bukowski}, H.~{Meijering},
  H.~{Meulman}, M.~{Mevius}, E.~{Middelberg}, R.~{Millenaar}, J.~C.~A.
  {Miller-Jones}, R.~N. {Mohan}, J.~D. {Mol}, J.~{Morawietz}, R.~{Morganti},
  D.~D. {Mulcahy}, E.~{Mulder}, H.~{Munk}, L.~{Nieuwenhuis}, R.~{van
  Nieuwpoort}, J.~E. {Noordam}, M.~{Norden}, A.~{Noutsos}, A.~R. {Offringa},
  H.~{Olofsson}, A.~{Omar}, E.~{Orr{\'u}}, R.~{Overeem}, H.~{Paas},
  M.~{Pandey-Pommier}, V.~N. {Pandey}, R.~{Pizzo}, A.~{Polatidis},
  D.~{Rafferty}, S.~{Rawlings}, W.~{Reich}, J.-P. {de Reijer}, J.~{Reitsma},
  G.~A. {Renting}, P.~{Riemers}, E.~{Rol}, J.~W. {Romein}, J.~{Roosjen},
  M.~{Ruiter}, A.~{Scaife}, K.~{van der Schaaf}, B.~{Scheers}, P.~{Schellart},
  A.~{Schoenmakers}, G.~{Schoonderbeek}, M.~{Serylak}, A.~{Shulevski},
  J.~{Sluman}, O.~{Smirnov}, C.~{Sobey}, H.~{Spreeuw}, M.~{Steinmetz}, C.~G.~M.
  {Sterks}, H.-J. {Stiepel}, K.~{Stuurwold}, M.~{Tagger}, Y.~{Tang},
  C.~{Tasse}, I.~{Thomas}, S.~{Thoudam}, M.~C. {Toribio}, B.~{van der Tol},
  O.~{Usov}, M.~{van Veelen}, A.-J. {van der Veen}, S.~{ter Veen}, J.~P.~W.
  {Verbiest}, R.~{Vermeulen}, N.~{Vermaas}, C.~{Vocks}, C.~{Vogt}, M.~{de Vos},
  E.~{van der Wal}, R.~{van Weeren}, H.~{Weggemans}, P.~{Weltevrede},
  S.~{White}, S.~J. {Wijnholds}, T.~{Wilhelmsson}, O.~{Wucknitz},
  S.~{Yatawatta}, P.~{Zarka}, A.~{Zensus}, and J.~{van Zwieten}.
\newblock {LOFAR: The LOw-Frequency ARray}.
\newblock {\em A\&A}, 556:A2, August 2013.

\bibitem{2013PASA...30....7T}
S.~J. {Tingay}, R.~{Goeke}, J.~D. {Bowman}, D.~{Emrich}, S.~M. {Ord}, D.~A.
  {Mitchell}, M.~F. {Morales}, T.~{Booler}, B.~{Crosse}, R.~B. {Wayth}, C.~J.
  {Lonsdale}, S.~{Tremblay}, D.~{Pallot}, T.~{Colegate}, A.~{Wicenec},
  N.~{Kudryavtseva}, W.~{Arcus}, D.~{Barnes}, G.~{Bernardi}, F.~{Briggs},
  S.~{Burns}, J.~D. {Bunton}, R.~J. {Cappallo}, B.~E. {Corey}, A.~{Deshpande},
  L.~{Desouza}, B.~M. {Gaensler}, L.~J. {Greenhill}, P.~J. {Hall}, B.~J.
  {Hazelton}, D.~{Herne}, J.~N. {Hewitt}, M.~{Johnston-Hollitt}, D.~L.
  {Kaplan}, J.~C. {Kasper}, B.~B. {Kincaid}, R.~{Koenig}, E.~{Kratzenberg},
  M.~J. {Lynch}, B.~{Mckinley}, S.~R. {Mcwhirter}, E.~{Morgan}, D.~{Oberoi},
  J.~{Pathikulangara}, T.~{Prabu}, R.~A. {Remillard}, A.~E.~E. {Rogers},
  A.~{Roshi}, J.~E. {Salah}, R.~J. {Sault}, N.~{Udaya-Shankar},
  F.~{Schlagenhaufer}, K.~S. {Srivani}, J.~{Stevens}, R.~{Subrahmanyan},
  M.~{Waterson}, R.~L. {Webster}, A.~R. {Whitney}, A.~{Williams}, C.~L.
  {Williams}, and J.~S.~B. {Wyithe}.
\newblock {The Murchison Widefield Array: The Square Kilometre Array Precursor
  at Low Radio Frequencies}.
\newblock {\em PASA}, 30:7, January 2013.

\bibitem{2010AJ....139.1468P}
A.~R. {Parsons}, D.~C. {Backer}, G.~S. {Foster}, M.~C.~H. {Wright}, R.~F.
  {Bradley}, N.~E. {Gugliucci}, C.~R. {Parashare}, E.~E. {Benoit}, J.~E.
  {Aguirre}, D.~C. {Jacobs}, C.~L. {Carilli}, D.~{Herne}, M.~J. {Lynch}, J.~R.
  {Manley}, and D.~J. {Werthimer}.
\newblock {The Precision Array for Probing the Epoch of Re-ionization: Eight
  Station Results}.
\newblock {\em AJ}, 139:1468--1480, April 2010.

\bibitem{2017PASP..129d5001D}
D.~R. {DeBoer}, A.~R. {Parsons}, J.~E. {Aguirre}, P.~{Alexander}, Z.~S. {Ali},
  A.~P. {Beardsley}, G.~{Bernardi}, J.~D. {Bowman}, R.~F. {Bradley}, C.~L.
  {Carilli}, C.~{Cheng}, E.~{de Lera Acedo}, J.~S. {Dillon}, A.~{Ewall-Wice},
  G.~{Fadana}, N.~{Fagnoni}, R.~{Fritz}, S.~R. {Furlanetto}, B.~{Glendenning},
  B.~{Greig}, J.~{Grobbelaar}, B.~J. {Hazelton}, J.~N. {Hewitt}, J.~{Hickish},
  D.~C. {Jacobs}, A.~{Julius}, M.~{Kariseb}, S.~A. {Kohn}, T.~{Lekalake},
  A.~{Liu}, A.~{Loots}, D.~{MacMahon}, L.~{Malan}, C.~{Malgas}, M.~{Maree},
  Z.~{Martinot}, N.~{Mathison}, E.~{Matsetela}, A.~{Mesinger}, M.~F. {Morales},
  A.~R. {Neben}, N.~{Patra}, S.~{Pieterse}, J.~C. {Pober}, N.~{Razavi-Ghods},
  J.~{Ringuette}, J.~{Robnett}, K.~{Rosie}, R.~{Sell}, C.~{Smith}, A.~{Syce},
  M.~{Tegmark}, N.~{Thyagarajan}, P.~K.~G. {Williams}, and H.~{Zheng}.
\newblock {Hydrogen Epoch of Reionization Array (HERA)}.
\newblock {\em PASP}, 129(4):045001, April 2017.

\bibitem{2012IJMPS..12..256C}
X.~{Chen}.
\newblock {The Tianlai Project: a 21CM Cosmology Experiment}.
\newblock In {\em International Journal of Modern Physics Conference Series},
  volume~12 of {\em International Journal of Modern Physics Conference Series},
  pages 256--263, March 2012.

\bibitem{2014SPIE.9145E..22B}
K.~{Bandura}, G.~E. {Addison}, M.~{Amiri}, J.~R. {Bond}, D.~{Campbell-Wilson},
  L.~{Connor}, J.-F. {Cliche}, G.~{Davis}, M.~{Deng}, N.~{Denman}, M.~{Dobbs},
  M.~{Fandino}, K.~{Gibbs}, A.~{Gilbert}, M.~{Halpern}, D.~{Hanna}, A.~D.
  {Hincks}, G.~{Hinshaw}, C.~{H{\"o}fer}, P.~{Klages}, T.~L. {Landecker},
  K.~{Masui}, J.~{Mena Parra}, L.~B. {Newburgh}, U.-l. {Pen}, J.~B. {Peterson},
  A.~{Recnik}, J.~R. {Shaw}, K.~{Sigurdson}, M.~{Sitwell}, G.~{Smecher},
  R.~{Smegal}, K.~{Vanderlinde}, and D.~{Wiebe}.
\newblock {Canadian Hydrogen Intensity Mapping Experiment (CHIME) pathfinder}.
\newblock In {\em Ground-based and Airborne Telescopes V}, volume 9145 of {\em
  procspie}, page 914522, July 2014.

\bibitem{2013arXiv1311.4288H}
M.~{Huynh} and J.~{Lazio}.
\newblock {An Overview of the Square Kilometre Array}.
\newblock {\em ArXiv:1311.4288}, November 2013.

\bibitem{2008MNRAS.389.1319J}
V.~{Jeli{\'c}}, S.~{Zaroubi}, P.~{Labropoulos}, R.~M. {Thomas}, G.~{Bernardi},
  M.~A. {Brentjens}, A.~G. {de Bruyn}, B.~{Ciardi}, G.~{Harker}, L.~V.~E.
  {Koopmans}, V.~N. {Pandey}, J.~{Schaye}, and S.~{Yatawatta}.
\newblock {Foreground simulations for the LOFAR-epoch of reionization
  experiment}.
\newblock {\em MNRAS}, 389:1319--1335, September 2008.

\bibitem{1999A&A...345..380S}
P.~A. {Shaver}, R.~A. {Windhorst}, P.~{Madau}, and A.~G. {de Bruyn}.
\newblock {Can the reionization epoch be detected as a global signature in the
  cosmic background?}
\newblock {\em A\&A}, 345:380--390, May 1999.

\bibitem{2004MNRAS.355.1053D}
T.~{Di Matteo}, B.~{Ciardi}, and F.~{Miniati}.
\newblock {The 21-cm emission from the reionization epoch: extended and point
  source foregrounds}.
\newblock {\em MNRAS}, 355:1053--1065, December 2004.

\bibitem{2008MNRAS.388..247D}
A.~{de Oliveira-Costa}, M.~{Tegmark}, B.~M. {Gaensler}, J.~{Jonas}, T.~L.
  {Landecker}, and P.~{Reich}.
\newblock {A model of diffuse Galactic radio emission from 10 MHz to 100 GHz}.
\newblock {\em MNRAS}, 388:247--260, July 2008.

\bibitem{2012MNRAS.419.3491L}
A.~{Liu} and M.~{Tegmark}.
\newblock {How well can we measure and understand foregrounds with 21-cm
  experiments?}
\newblock {\em MNRAS}, 419:3491--3504, February 2012.

\bibitem{2017MNRAS.464.3486Z}
H.~{Zheng}, M.~{Tegmark}, J.~S. {Dillon}, D.~A. {Kim}, A.~{Liu}, A.~R. {Neben},
  J.~{Jonas}, P.~{Reich}, and W.~{Reich}.
\newblock {An improved model of diffuse galactic radio emission from 10 MHz to
  5 THz}.
\newblock {\em MNRAS}, 464:3486--3497, January 2017.

\bibitem{2014ApJ...781...57S}
J.~R. {Shaw}, K.~{Sigurdson}, U.-L. {Pen}, A.~{Stebbins}, and M.~{Sitwell}.
\newblock {All-sky Interferometry with Spherical Harmonic Transit Telescopes}.
\newblock {\em ApJ}, 781:57, February 2014.

\bibitem{2015PhRvD..91h3514S}
J.~R. {Shaw}, K.~{Sigurdson}, M.~{Sitwell}, A.~{Stebbins}, and U.-L. {Pen}.
\newblock {Coaxing cosmic 21 cm fluctuations from the polarized sky using m
  -mode analysis}.
\newblock {\em Phys. Rev.~D}, 91(8):083514, April 2015.

\bibitem{1982AandAS...47....1H}
C.~G.~T. {Haslam}, C.~J. {Salter}, H.~{Stoffel}, and W.~E. {Wilson}.
\newblock {A 408 MHz all-sky continuum survey. II - The atlas of contour maps}.
\newblock {\em A\&AS}, 47:1, January 1982.

\bibitem{2007AJ....134.1245C}
A.~S. {Cohen}, W.~M. {Lane}, W.~D. {Cotton}, N.~E. {Kassim}, T.~J.~W. {Lazio},
  R.~A. {Perley}, J.~J. {Condon}, and W.~C. {Erickson}.
\newblock {The VLA Low-Frequency Sky Survey}.
\newblock {\em AJ}, 134:1245--1262, September 2007.

\bibitem{1998AJ....115.1693C}
J.~J. {Condon}, W.~D. {Cotton}, E.~W. {Greisen}, Q.~F. {Yin}, R.~A. {Perley},
  G.~B. {Taylor}, and J.~J. {Broderick}.
\newblock {The NRAO VLA Sky Survey}.
\newblock {\em AJ}, 115:1693--1716, May 1998.

\bibitem{2002ApJ...564..576D}
T.~{Di Matteo}, R.~{Perna}, T.~{Abel}, and M.~J. {Rees}.
\newblock {Radio Foregrounds for the 21 Centimeter Tomography of the Neutral
  Intergalactic Medium at High Redshifts}.
\newblock {\em ApJ}, 564:576--580, January 2002.

\bibitem{2005ApJ...625..575S}
M.~G. {Santos}, A.~{Cooray}, and L.~{Knox}.
\newblock {Multifrequency Analysis of 21 Centimeter Fluctuations from the Era
  of Reionization}.
\newblock {\em ApJ}, 625:575--587, June 2005.

\bibitem{2018arXiv180505222B}
A.~{Bonaldi}, M.~{Bonato}, V.~{Galluzzi}, I.~{Harrison}, M.~{Massardi},
  S.~{Kay}, G.~{De Zotti}, and M.~L. {Brown}.
\newblock {The Tiered Radio Extragalactic Continuum Simulation (T-RECS)}.
\newblock {\em ArXiv:1805.05222}, May 2018.

\bibitem{2015MNRAS.451.4311R}
M.~{Remazeilles}, C.~{Dickinson}, A.~J. {Banday}, M.-A. {Bigot-Sazy}, and
  T.~{Ghosh}.
\newblock {An improved source-subtracted and destriped 408-MHz all-sky map}.
\newblock {\em MNRAS}, 451:4311--4327, August 2015.

\bibitem{2001MNRAS.325.1241V}
P.~{V{\"a}is{\"a}nen}, E.~V. {Tollestrup}, and G.~G. {Fazio}.
\newblock {Confusion limit resulting from galaxies: using the Infrared Array
  Camera on board SIRTF}.
\newblock {\em MNRAS}, 325:1241--1252, August 2001.

\bibitem{2006MNRAS.369..281J}
W.-S. {Jeong}, C.~P. {Pearson}, H.~M. {Lee}, S.~{Pak}, and T.~{Nakagawa}.
\newblock {Far-infrared detection limits - II. Probing confusion including
  source confusion}.
\newblock {\em MNRAS}, 369:281--294, June 2006.

\bibitem{2012ApJ...758...23C}
J.~J. {Condon}, W.~D. {Cotton}, E.~B. {Fomalont}, K.~I. {Kellermann},
  N.~{Miller}, R.~A. {Perley}, D.~{Scott}, T.~{Vernstrom}, and J.~V. {Wall}.
\newblock {Resolving the Radio Source Background: Deeper Understanding through
  Confusion}.
\newblock {\em ApJ}, 758:23, October 2012.

\bibitem{1973ApJ...183....1M}
H.~S. {Murdoch}, D.~F. {Crawford}, and D.~L. {Jauncey}.
\newblock {Maximum-Likelihood Estimation of the Number-Flux Distribution of
  Radio Sources in the Presence of Noise and Confusion}.
\newblock {\em ApJ}, 183:1--14, July 1973.

\bibitem{2005ApJ...622..759G}
K.~M. {G{\'o}rski}, E.~{Hivon}, A.~J. {Banday}, B.~D. {Wandelt}, F.~K.
  {Hansen}, M.~{Reinecke}, and M.~{Bartelmann}.
\newblock {HEALPix: A Framework for High-Resolution Discretization and Fast
  Analysis of Data Distributed on the Sphere}.
\newblock {\em ApJ}, 622:759--771, April 2005.

\bibitem{1976MNRAS.177..601C}
J.~L. {Caswell}.
\newblock {A map of the northern sky at 10 MHz}.
\newblock {\em MNRAS}, 177:601--616, December 1976.

\bibitem{1999AandAS..137....7R}
R.~S. {Roger}, C.~H. {Costain}, T.~L. {Landecker}, and C.~M. {Swerdlyk}.
\newblock {The radio emission from the Galaxy at 22 MHz}.
\newblock {\em A\&AS}, 137:7--19, May 1999.

\bibitem{2011AandA...525A.138G}
A.~E. {Guzm{\'a}n}, J.~{May}, H.~{Alvarez}, and K.~{Maeda}.
\newblock {All-sky Galactic radiation at 45 MHz and spectral index between 45
  and 408 MHz}.
\newblock {\em A\&A}, 525:A138, January 2011.

\bibitem{1970AuJPA..16....1L}
T.~L. {Landecker} and R.~{Wielebinski}.
\newblock {The Galactic Metre Wave Radiation: A two-frequency survey between
  declinations +25$^{o}$ and -25$^{o}$ and the preparation of a map of the
  whole sky}.
\newblock {\em Australian Journal of Physics Astrophysical Supplement}, 16:1,
  1970.

\bibitem{2001AandA...376..861R}
P.~{Reich}, J.~C. {Testori}, and W.~{Reich}.
\newblock {A radio continuum survey of the southern sky at 1420 MHz. The atlas
  of contour maps}.
\newblock {\em A\&A}, 376:861--877, September 2001.

\bibitem{2013yCat..35560001T}
C.~{Tello}, T.~{Villela}, S.~{Torres}, M.~{Bersanelli}, G.~F. {Smoot}, I.~S.
  {Ferreira}, A.~{Cingoz}, J.~{Lamb}, D.~{Barbosa}, D.~{Perez-Becker},
  S.~{Ricciardi}, J.~A. {Currivan}, P.~{Platania}, and D.~{Maino}.
\newblock {VizieR Online Data Catalog: The 2.3GHz continuum survey of the GEM
  project (Tello+, 2013)}.
\newblock {\em VizieR Online Data Catalog}, 355, May 2013.

\bibitem{2013ApJS..208...20B}
C.~L. {Bennett}, D.~{Larson}, J.~L. {Weiland}, N.~{Jarosik}, G.~{Hinshaw},
  N.~{Odegard}, K.~M. {Smith}, R.~S. {Hill}, B.~{Gold}, M.~{Halpern},
  E.~{Komatsu}, M.~R. {Nolta}, L.~{Page}, D.~N. {Spergel}, E.~{Wollack},
  J.~{Dunkley}, A.~{Kogut}, M.~{Limon}, S.~S. {Meyer}, G.~S. {Tucker}, and
  E.~L. {Wright}.
\newblock {Nine-year Wilkinson Microwave Anisotropy Probe (WMAP) Observations:
  Final Maps and Results}.
\newblock {\em ApJS}, 208:20, October 2013.

\bibitem{2007ARNPS..57..285S}
A.~W. {Strong}, I.~V. {Moskalenko}, and V.~S. {Ptuskin}.
\newblock {Cosmic-Ray Propagation and Interactions in the Galaxy}.
\newblock {\em Annual Review of Nuclear and Particle Science}, 57:285--327,
  November 2007.

\bibitem{PhysRevD.82.092004}
M.~Ackermann, M.~Ajello, W.~B. Atwood, L.~Baldini, J.~Ballet, G.~Barbiellini,
  D.~Bastieri, B.~M. Baughman, K.~Bechtol, and F.~Bellardi.
\newblock Fermi lat observations of cosmic-ray electrons from 7 gev to 1 tev.
\newblock {\em Phys. Rev. D}, 82:092004, Nov 2010.

\bibitem{2011A&A...534A..54S}
A.~W. {Strong}, E.~{Orlando}, and T.~R. {Jaffe}.
\newblock {The interstellar cosmic-ray electron spectrum from synchrotron
  radiation and direct measurements}.
\newblock {\em A\&A}, 534:A54, October 2011.

\bibitem{1998astro.ph..1121S}
G.~F. {Smoot}.
\newblock {Galactic Free-free and H-alpha Emission}.
\newblock {\em ArXiv:astro-ph/9801121}, January 1998.

\bibitem{1998ApJ...505..473P}
P.~{Platania}, M.~{Bensadoun}, M.~{Bersanelli}, G.~{De Amici}, A.~{Kogut},
  S.~{Levin}, D.~{Maino}, and G.~F. {Smoot}.
\newblock {A Determination of the Spectral Index of Galactic Synchrotron
  Emission in the 1-10 GHz Range}.
\newblock {\em ApJ}, 505:473--483, October 1998.

\bibitem{1988A&AS...74....7R}
P.~{Reich} and W.~{Reich}.
\newblock {A map of spectral indices of the Galactic radio continuum emission
  between 408 MHz and 1420 MHz for the entire northern sky}.
\newblock {\em A\&AS}, 74:7--20, July 1988.

\bibitem{2001A&A...368.1123T}
J.~C. {Testori}, P.~{Reich}, J.~A. {Bava}, F.~R. {Colomb}, E.~E. {Hurrel},
  J.~J. {Larrarte}, W.~{Reich}, and A.~J. {Sanz}.
\newblock {A radio continuum survey of the southern sky at 1420 MHz.
  Observations and data reduction}.
\newblock {\em A\&A}, 368:1123--1132, March 2001.

\bibitem{1967MNRAS.136..219B}
A.~H. {Bridle}.
\newblock {The spectrum of the radio background between 13 and 404 MHz}.
\newblock {\em MNRAS}, 136:219, 1967.

\bibitem{1974MNRAS.166..345S}
G.~{Sironi}.
\newblock {The spectrum of the galactic non-thermal background radiation-1.
  Observations at 151-5 and 408 MHz}.
\newblock {\em MNRAS}, 166:345--354, February 1974.

\bibitem{1974MNRAS.166..355W}
A.~S. {Webster}.
\newblock {The spectrum of the galactic non-thermal background radiational
  Observations at 408, 610 and 1407 MHz}.
\newblock {\em MNRAS}, 166:355--372, February 1974.

\bibitem{1979MNRAS.189..465C}
H.~V. {Cane}.
\newblock {Spectra of the non-thermal radio radiation from the galactic polar
  regions}.
\newblock {\em MNRAS}, 189:465--478, November 1979.

\bibitem{1987MNRAS.225..307L}
K.~D. {Lawson}, C.~J. {Mayer}, J.~L. {Osborne}, and M.~L. {Parkinson}.
\newblock {Variations in the Spectral Index of the Galactic Radio Continuum
  Emission in the Northern Hemisphere}.
\newblock {\em MNRAS}, 225:307, March 1987.

\bibitem{1991MNRAS.248..705B}
A.~J. {Banday} and A.~W. {Wolfendale}.
\newblock {Fluctuations in the galactic synchrotron radiation. I - Implications
  for searches for fluctuations of cosmological origin}.
\newblock {\em MNRAS}, 248:705--714, February 1991.

\bibitem{2016A&A...594A...9P}
{Planck Collaboration}, R.~{Adam}, P.~A.~R. {Ade}, N.~{Aghanim}, M.~{Arnaud},
  M.~{Ashdown}, J.~{Aumont}, C.~{Baccigalupi}, A.~J. {Banday}, R.~B.
  {Barreiro}, and et~al.
\newblock {Planck 2015 results. IX. Diffuse component separation: CMB maps}.
\newblock {\em A\&A}, 594:A9, September 2016.

\bibitem{1984ApJ...284...44C}
J.~J. {Condon}.
\newblock {Cosmological evolution of radio sources found at 1.4 GHz}.
\newblock {\em ApJ}, 284:44--53, September 1984.

\bibitem{2005PASA...22...36J}
C.~{Jackson}.
\newblock {The Extragalactic Radio Sky at Faint Flux Densities}.
\newblock {\em pasa}, 22:36--48, 2005.

\bibitem{2016ApJ...831..168K}
K.~I. {Kellermann}, J.~J. {Condon}, A.~E. {Kimball}, R.~A. {Perley}, and {\v
  Z}.~{Ivezi{\'c}}.
\newblock {Radio-loud and Radio-quiet QSOs}.
\newblock {\em ApJ}, 831:168, November 2016.

\bibitem{2017ApJ...842...95M}
C.~{Mancuso}, A.~{Lapi}, I.~{Prandoni}, I.~{Obi}, J.~{Gonzalez-Nuevo},
  F.~{Perrotta}, A.~{Bressan}, A.~{Celotti}, and L.~{Danese}.
\newblock {Galaxy Evolution in the Radio Band: The Role of Star-forming
  Galaxies and Active Galactic Nuclei}.
\newblock {\em ApJ}, 842:95, June 2017.

\bibitem{1999AJ....117.1578B}
D.~C.-J. {Bock}, M.~I. {Large}, and E.~M. {Sadler}.
\newblock {SUMSS: A Wide-Field Radio Imaging Survey of the Southern Sky. I.
  Science Goals, Survey Design, and Instrumentation}.
\newblock {\em AJ}, 117:1578--1593, March 1999.

\bibitem{2003MNRAS.342.1117M}
T.~{Mauch}, T.~{Murphy}, H.~J. {Buttery}, J.~{Curran}, R.~W. {Hunstead},
  B.~{Piestrzynski}, J.~G. {Robertson}, and E.~M. {Sadler}.
\newblock {SUMSS: a wide-field radio imaging survey of the southern sky - II.
  The source catalogue}.
\newblock {\em MNRAS}, 342:1117--1130, July 2003.

\bibitem{1959MmRAS..68...37E}
D.~O. {Edge}, J.~R. {Shakeshaft}, W.~B. {McAdam}, J.~E. {Baldwin}, and
  S.~{Archer}.
\newblock {A survey of radio sources at a frequency of 159 Mc/s.}
\newblock {\em Mem.~RAS}, 68:37--60, 1959.

\bibitem{1975MNRAS.171..475P}
T.~J. {Pearson}.
\newblock {The 5C 5 survey of radio sources}.
\newblock {\em MNRAS}, 171:475--505, June 1975.

\bibitem{1995ApJ...450..559B}
R.~H. {Becker}, R.~L. {White}, and D.~J. {Helfand}.
\newblock {The FIRST Survey: Faint Images of the Radio Sky at Twenty
  Centimeters}.
\newblock {\em ApJ}, 450:559, September 1995.

\bibitem{2015ApJ...801...26H}
D.~J. {Helfand}, R.~L. {White}, and R.~H. {Becker}.
\newblock {The Last of FIRST: The Final Catalog and Source Identifications}.
\newblock {\em ApJ}, 801:26, March 2015.

\bibitem{2003A&A...405...53O}
R.~A. {Overzier}, H.~J.~A. {R{\"o}ttgering}, R.~B. {Rengelink}, and R.~J.
  {Wilman}.
\newblock {The spatial clustering of radio sources in NVSS and FIRST;
  implications for galaxy clustering evolution}.
\newblock {\em A\&A}, 405:53--72, July 2003.

\bibitem{1975CRASM.280.1551M}
B.~{Mandelbrot}.
\newblock {On a decomposable model of a hierarchical universe - Derivation of
  galactic correlations on the celestial sphere}.
\newblock {\em Academie des Sciences Paris Comptes Rendus Serie Sciences
  Mathematiques}, 280:1551--1554, June 1975.

\bibitem{1995ApJS...96..401C}
S.~{Colombi}, F.~R. {Bouchet}, and R.~{Schaeffer}.
\newblock {A count probability cookbok: Spurious effects and the scaling
  model}.
\newblock {\em ApJS}, 96:401--428, February 1995.

\bibitem{2003AJ....125.2064G}
R.~R. {Gal}, R.~R. {de Carvalho}, P.~A.~A. {Lopes}, S.~G. {Djorgovski}, R.~J.
  {Brunner}, A.~{Mahabal}, and S.~C. {Odewahn}.
\newblock {The Northern Sky Optical Cluster Survey. II. An Objective Cluster
  Catalog for 5800 Square Degrees}.
\newblock {\em AJ}, 125:2064--2084, April 2003.

\bibitem{2014MNRAS.440...10A}
D.~{Alonso}, A.~{Bueno Belloso}, F.~J. {S{\'a}nchez},
  J.~{Garc{\'{\i}}a-Bellido}, and E.~{S{\'a}nchez}.
\newblock {Measuring the transition to homogeneity with photometric redshift
  surveys}.
\newblock {\em MNRAS}, 440:10--23, May 2014.

\bibitem{2018A&A...613A..58V}
C.~L. {Van Eck}, M.~{Haverkorn}, M.~I.~R. {Alves}, R.~{Beck}, P.~{Best},
  E.~{Carretti}, K.~T. {Chy{\.z}y}, J.~S. {Farnes}, K.~{Ferri{\`e}re}, M.~J.
  {Hardcastle}, G.~{Heald}, C.~{Horellou}, M.~{Iacobelli}, V.~{Jeli{\'c}},
  D.~D. {Mulcahy}, S.~P. {O'Sullivan}, I.~M. {Polderman}, W.~{Reich}, C.~J.
  {Riseley}, H.~{R{\"o}ttgering}, D.~H.~F.~M. {Schnitzeler}, T.~W. {Shimwell},
  V.~{Vacca}, J.~{Vink}, and G.~J. {White}.
\newblock {Polarized point sources in the LOFAR Two-meter Sky Survey: A
  preliminary catalog}.
\newblock {\em A\&A}, 613:A58, June 2018.

\bibitem{2010MNRAS.409.1647J}
V.~{Jeli{\'c}}, S.~{Zaroubi}, P.~{Labropoulos}, G.~{Bernardi}, A.~G. {de
  Bruyn}, and L.~V.~E. {Koopmans}.
\newblock {Realistic simulations of the Galactic polarized foreground:
  consequences for 21-cm reionization detection experiments}.
\newblock {\em MNRAS}, 409:1647--1659, December 2010.

\bibitem{Asad:2015sda}
Khan M.~B. Asad et~al.
\newblock {Polarization leakage in epoch of reionization windows ? I. Low
  Frequency Array observations of the 3C196 field}.
\newblock {\em Mon. Not. Roy. Astron. Soc.}, 451(4):3709--3727, 2015.

\bibitem{2016MNRAS.462.4482A}
K.~M.~B. {Asad}, L.~V.~E. {Koopmans}, V.~{Jeli{\'c}}, A.~{Ghosh}, F.~B.
  {Abdalla}, M.~A. {Brentjens}, A.~G. {de Bruyn}, B.~{Ciardi}, B.~K. {Gehlot},
  I.~T. {Iliev}, M.~{Mevius}, V.~N. {Pandey}, S.~{Yatawatta}, and S.~{Zaroubi}.
\newblock {Polarization leakage in epoch of reionization windows - II. Primary
  beam model and direction-dependent calibration}.
\newblock {\em MNRAS}, 462:4482--4494, November 2016.

\bibitem{2017A&A...597A..98V}
C.~L. {Van Eck}, M.~{Haverkorn}, M.~I.~R. {Alves}, R.~{Beck}, A.~G. {de Bruyn},
  T.~{En{\ss}lin}, J.~S. {Farnes}, K.~{Ferri{\`e}re}, G.~{Heald},
  C.~{Horellou}, A.~{Horneffer}, M.~{Iacobelli}, V.~{Jeli{\'c}},
  I.~{Mart{\'{\i}}-Vidal}, D.~D. {Mulcahy}, W.~{Reich}, H.~J.~A.
  {R{\"o}ttgering}, A.~M.~M. {Scaife}, D.~H.~F.~M. {Schnitzeler}, C.~{Sobey},
  and S.~S. {Sridhar}.
\newblock {Faraday tomography of the local interstellar medium with LOFAR:
  Galactic foregrounds towards IC 342}.
\newblock {\em A\&A}, 597:A98, January 2017.

\bibitem{2018MNRAS.476.3051A}
K.~M.~B. {Asad}, L.~V.~E. {Koopmans}, V.~{Jeli{\'c}}, A.~G. {de Bruyn}, V.~N.
  {Pandey}, and B.~K. {Gehlot}.
\newblock {Polarization leakage in epoch of reionization windows - III.
  Wide-field effects of narrow-field arrays}.
\newblock {\em MNRAS}, 476:3051--3062, May 2018.

\bibitem{Spinelli:2018kpm}
M.~Spinelli, G.~Bernardi, and M.~G. Santos.
\newblock {Simulations of Galactic polarized synchrotron emission for Epoch of
  Reionization observations}.
\newblock 2018.

\bibitem{1987A&A...184....7T}
L.~{Toffolatti}, A.~{Franceschini}, L.~{Danese}, and G.~{de Zotti}.
\newblock {The local radio luminosity function of galaxies}.
\newblock {\em A\&A}, 184:7--15, October 1987.

\bibitem{1979ApJ...232..352S}
A.~{Sandage}, G.~A. {Tammann}, and A.~{Yahil}.
\newblock {The velocity field of bright nearby galaxies. I - The variation of
  mean absolute magnitude with redshift for galaxies in a magnitude-limited
  sample}.
\newblock {\em ApJ}, 232:352--364, September 1979.

\bibitem{2007MNRAS.375..931M}
T.~{Mauch} and E.~M. {Sadler}.
\newblock {Radio sources in the 6dFGS: local luminosity functions at 1.4GHz for
  star-forming galaxies and radio-loud AGN}.
\newblock {\em MNRAS}, 375:931--950, March 2007.

\bibitem{2016A&A...594A..13P}
{Planck Collaboration}, P.~A.~R. {Ade}, N.~{Aghanim}, M.~{Arnaud},
  M.~{Ashdown}, J.~{Aumont}, C.~{Baccigalupi}, A.~J. {Banday}, R.~B.
  {Barreiro}, J.~G. {Bartlett}, and et~al.
\newblock {Planck 2015 results. XIII. Cosmological parameters}.
\newblock {\em A\&A}, 594:A13, September 2016.

\bibitem{1990MNRAS.242..318S}
W.~{Saunders}, M.~{Rowan-Robinson}, A.~{Lawrence}, G.~{Efstathiou},
  N.~{Kaiser}, R.~S. {Ellis}, and C.~S. {Frenk}.
\newblock {The 60-micron and far-infrared luminosity functions of IRAS
  galaxies}.
\newblock {\em MNRAS}, 242:318--337, January 1990.

\bibitem{1989ApJ...338...13C}
J.~J. {Condon}.
\newblock {The 1.4 gigahertz luminosity function and its evolution}.
\newblock {\em ApJ}, 338:13--23, March 1989.

\end{thebibliography}
\end{multicols}
\end{document}